\documentclass[reprint,twocoloumn,superscriptaddress,floatfix]{revtex4-2}
\usepackage[utf8]{inputenc}
\usepackage{listings} 
\usepackage[dvipsnames]{xcolor}
\usepackage{amssymb,amsmath,amsbsy,mathtools,nccmath}
\DeclareMathOperator{\sgn}{sgn}
\usepackage{txfonts}
\usepackage{lmodern}
\usepackage{xfrac}
\usepackage{graphicx}
\usepackage{booktabs}
\usepackage{cellspace}
\usepackage{caption}
\usepackage{subcaption}
\usepackage{comment}
\usepackage{svg}
\newcommand{\lmfp}{\ensuremath{\lambda_{mf}}}

\begin{document}

\title{Unified Model of Heated Plasma Expansion}
\date{\today}
\author{Ritwik Sain}
\affiliation{Center for High Energy Density Science, The University of Texas at Austin, Austin, TX 78712}
\author{Lance Labun}
\affiliation{Center for High Energy Density Science, The University of Texas at Austin, Austin, TX 78712}
\affiliation{Tau Systems Inc,  Austin, Texas, 78701, USA}
\author{Ou Z. Labun}
\affiliation{Center for High Energy Density Science, The University of Texas at Austin, Austin, TX 78712}
\author{Bjorn Manuel Hegelich}
\affiliation{Center for High Energy Density Science, The University of Texas at Austin, Austin, TX 78712}
\affiliation{Tau Systems Inc,  Austin, Texas, 78701, USA}

\begin{abstract}
       Motivated by the need to predict plasma density and temperature distributions created in the early stages of high-intensity laser-plasma interactions, we develop a fluid model of plasma expansion into vacuum that incorporates external heating.
    We propose a new three-parameter family of self-similar solutions for plasma expansion that models a wide range of spatiotemporal variations of the electron temperature. Depending on the relative scales of the heated plasma domain $L$, the Debye length $\lambda_D$ and an emergent ion-acoustic correlation length $\lambda_s$, characterized by the parameters $\frac{\lambda_s}{\lambda_D}$ and $\frac{L}{\lambda_s}$, a spectrum of dynamical behaviors for the expanding plasma are identified. The behavior is classified into five dynamical regimes, ranging from nearly quasineutral expansion to the formation of bare ion slabs susceptible to Coulomb explosion. Self-similar solutions with spatially uniform electron temperature with exponential time dependence are analyzed, and the dynamics in the five asymptotic limits in the parameter space are detailed. Scaling relations for the length scales and energies of the expanding plasma are proposed. The self-similar framework is applied to laser–plasma interactions, specifically addressing the plasma dynamics at a target surface as the target interacts with the rising laser intensity envelope. The results offer insights into the expansion behavior based on the laser-plasma parameters, and scaling relations for optimizing laser-plasma schemes and guiding experimental designs in high-intensity laser experiments.
\end{abstract}

\maketitle

\section{Introduction}
Plasma expansion into vacuum plays a critical role in many milestone achievements of laser-plasma accelerators: thin, solid-density targets generating sheath fields that accelerate protons to $\sim 100 ~\rm MeV$~\cite{higginson2018near} and $\text{C}^{6+}$ ions to $\sim 185 ~ \rm MeV$~\cite{henig2009enhanced}, and nanoparticle-plasmas assisting laser wakefield acceleration of electrons to $\sim 10 ~ \rm GeV$~\cite{aniculaesei2024acceleration}. In such experiments, the ionization threshold intensity can be separated from the peak laser intensity by several picoseconds to nanoseconds during which the laser may continue heating the plasma, raising the question how much the initially solid-density plasma expands. As the correct description of these experiments depends on the density and temperature distributions of the plasma leading up to the interaction of the peak laser intensity with the plasma, an accurate model of the expansion including heating would greatly enhance our understanding and control of the plasma conditions for acceleration. 

Plasma expansion into vacuum occurs in diverse contexts beyond laser-plasma accelerators, including inertial confinement fusion~\cite{Craxton_DirDriveICFRev_2015,Brueckner_ICFRev_1974}, nanoscale plasmas~\cite{SAXENA_HydroNanoPExp_2015}, plasmoids in fusion devices~\cite{Parks_PelletAbl_1978,MiloraRev_FusionPellet_1995}, space and astrophysical plasmas~\cite{Banks_PolarPlasma_1968,Borovsky_SupernoRem_1984,Moslem_WhiteDwarfExp_2012,Pantellini_SpaceDustExp_2012}. Other laser-plasma interaction applications include ion beam generation~\cite{MoraSelfSimilar2003,Roth_TNSA_2017}, vacuum acceleration of electrons~\cite{Thevenet_VacAcc_2016}, high-harmonic generation~\cite{Bulanov_HighHarm_1994}, relativistic induced transparency~\cite{Yin_BOAnsTarget_2011}, and coherent wake emission~\cite{Borot_CohWakeEmm_2012}. For the present paper though, we will be guided by physics and scales of laser-plasma experiments, looking to broader applicability of the model in future work.

Two regimes of expansion are well-known by their qualitative features: $\rm (1)$ Coulomb explosion, in which the electrons disperse much faster than the ion response time, leaving behind a highly-charged pre-dominantly-ion plasma that converts its electrostatic potential energy into ion kinetic energy, and $\rm (2)$ quasineutral expansion, in which the electron and ion densities remain tightly coupled and the expansion is driven by smaller scale electrostatic fields, especially at the vacuum interface where the lighter electrons lead the ions by a Debye length or so. The Debye length $\lambda_D=\left(\frac{\epsilon_0 T_{e}}{n_{e}e^2}\right)^{\sfrac{1}{2}}$ appears as the length scale of electrostatic fields created by charge separation between hotter, more mobile electrons and colder, slower ions. The two regimes could be characterized as opposite limits of the ratio $\sfrac{L}{\lambda_D}$, where $L$ is the characteristic length scale of the plasma. Plasmas leading to Coulomb explosion have $\lambda_D\gg L$, whereas quasineutral expansion has $\lambda_D\ll L$.

\begin{figure}[!hbtp]
    \centering
    \includegraphics[width=1\linewidth]{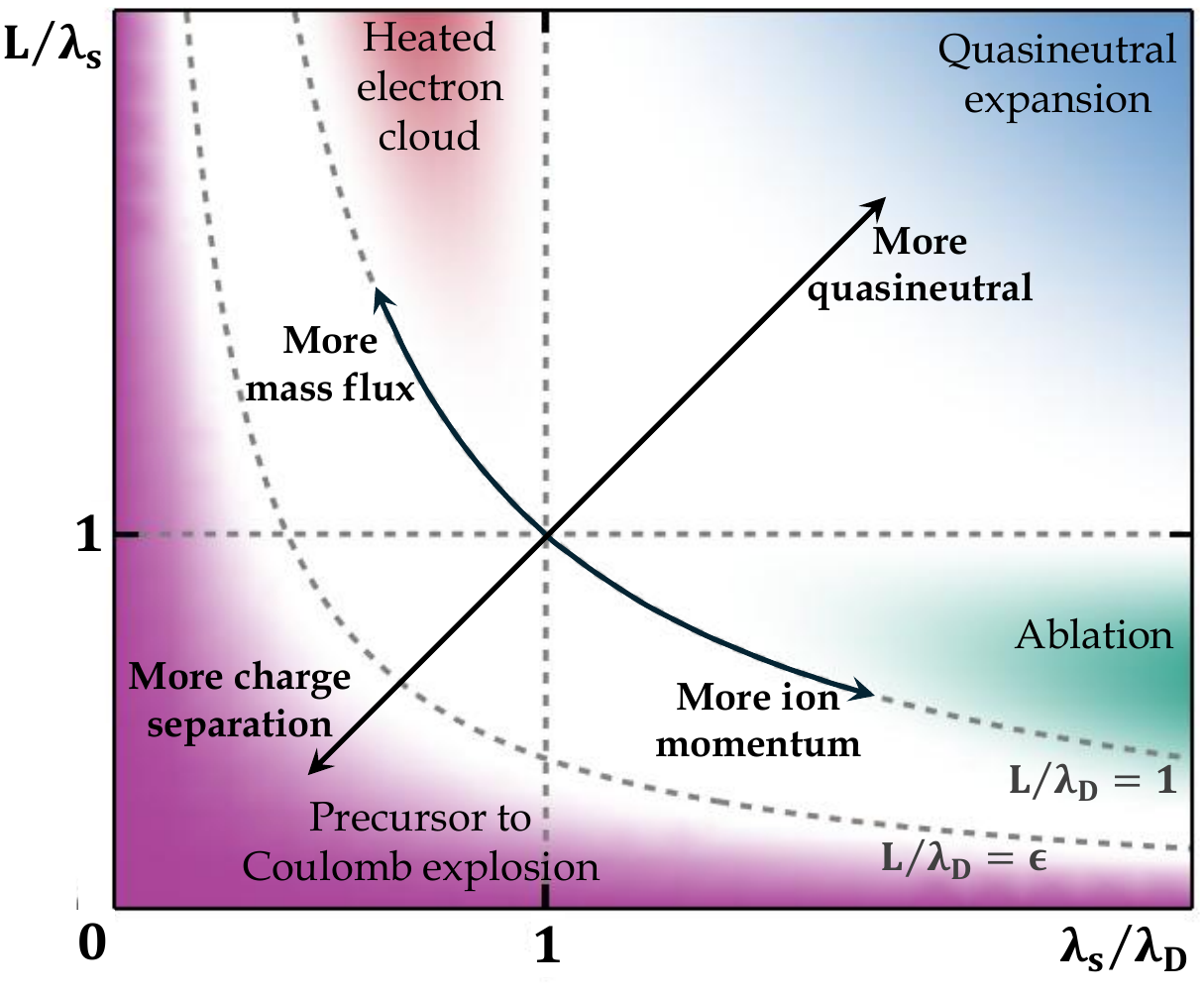}
    \caption{Parameter space of expansion phenomena characterized by $\sfrac{L}{\lambda_s}$ and $\sfrac{\lambda_s}{\lambda_D}$. The shaded regions represent the limits of Coulomb explosion precursor, quasineutral expansion, ablation, and expanding hot electron cloud}
    \label{fig:regSchematic}
\end{figure}

The inclusion of continuous heating on a timescale similar to the expansion time $\gamma^{-1}$ adds a new dimension to the phenomenology. To continue working in terms of spatial scales, we consider the length scale for ion-acoustic perturbations $\lambda_s=\sfrac{C_s}{\gamma}$ by combining this timescale with the ion speed of sound $C_s=(\sfrac{ZT_e}{m_i})^{\sfrac{1}{2}}$.  We will find that $\lambda_s$ splits the ratio $\sfrac{L}{\lambda_D}$ into two new dimensionless parameters, $\sfrac{L}{\lambda_s}$ and $\sfrac{\lambda_s}{\lambda_D}$ which characterize the parameter space of possible expansion phenomena as shown in Fig.\ref{fig:regSchematic}. On these axes, fixed values of the ratio $\sfrac{L}{\lambda_D}$ are hyperbolas, so that the Coulomb explosion regime $\sfrac{L}{\lambda_D}\ll1$ is found in the lower left, close to both axes, while the quasineutral regime $\sfrac{L}{\lambda_D}\gg 1$ is found in the upper right.  

Our purpose is to describe this two-dimensional parameter space of expansion physics. We derive a model for plasma expansion into vacuum consistently incorporating external heating that unifies the description of plasmas that lead to Coulomb explosion, quasineutral expansion, as well as two more limiting regimes that we might describe as expanding ablation-like and ``hot electron cloud'' expansion, as sketched in Fig.~\ref{fig:regSchematic}. The unified model covers all possible orderings of the length scales $L$, $\lambda_D$ and $\lambda_s$. The two new regimes represent the limits on the ratio of the plasma mass flow rate to the characteristic momentum density gained by the accelerated ions. When the mass flow rate is low, the ions gain energy and redistribute their momentum much faster than mass accretes into the expanding plasma. This leads to ablation near the vacuum interface, where diffuse ions with a high energy are rapidly produced. In the opposite limit, hot electron mass flux leads to an expanding hot electron cloud near the interface, while the ions do not respond appreciably in this time.

In laser-plasma experiments, the plasma is generated through field ionization of a target surface by the laser, and the free electrons are accelerated and heated thousands of times faster than ions due to the lower mass of the electrons. The hotter electrons diffuse, creating a sheath around the ions near the vacuum-target interface, and setting up an electrostatic field that accelerates the ions towards the vacuum~\cite{SACK1987311,Crow_Auer_Allen_1975}. The sheath plays a central role in the expansion dynamics by setting the scales of the electron density variation and charge separation, thus controlling the degree of quasineutrality. Moreover, being the driver for the expanding ions, the sheath field also governs the velocity scale of the accelerated ions and the energy partitioning among electrons, ions and the field. 

As the ions begin to move under the influence of this electrostatic field, we recognize three regions from vacuum side to bulk plasma side: $\rm (1)$ the sheath, where ion density is negligible, $\rm (2)$ the dynamically expanding plasma, where ions and electrons interact with each other while flowing with net velocity toward the vacuum, and $\rm (3)$ the unperturbed plasma, where the ion/plasma density remains close to its initial value.  The collisional mean free path quickly exceeds the length scale of the expansion region, and its strong dependence on temperature and density $\lambda_{\rm mf}\sim T_e^2n_e^{-1}$ ensures that the plasma typically remains collisionless during the heated expansion.

Following this picture and previous models, we employ the non-relativistic, collisionless fluid equations, which are valid from shortly after ionization until the laser intensity much exceeds $\sim 10^{18} ~\rm \sfrac{W}{cm^2}$ or the electron temperature becomes relativistic. Resolving the electron sheath requires writing the fluid equations separately for the ion and electron fluids and coupling them through the electrostatic field as determined from the net charge density by the Poisson equation~\cite{MoraSelfSimilar2003,MurakamiSelfSimilar2006,Beck_SelfSim_2009}.  We therefore start with a coupled set of partial differential equations (PDEs) for the electron and ion densities and velocities and the electrostatic field. With the ions assumed to remain cold (effectively zero temperature) throughout, the last degree of freedom to be determined is the electron temperature. The rate of laser-heating of the electrons depends on the electron density and temperature and the laser intensity (itself a function of time), $\frac{dT_e}{dt} = Q(n_e,T_e,t)$, with $Q$ either completely prescribed or dynamically determined by the solution through a closure relation. As a step toward control and applications, we seek the parametric dependence of the expansion dynamics on the laser and plasma parameters. Moreover, we would like to capture universal behavior, rather than dependence on details of the initial conditions such as laser temporal profile or initial plasma profile. These goals recommend finding and studying self-similar solutions to the fluid equations.

Self-similar solutions have been used to model plasma expansion in a wide range of plasmas: laser-produced plasmas~\cite{Anderson_SSwithBfield_1980,Murukami_LaserIrradiatedExp_2005,BOUDESOCQUE_SSAblFlowICF_2008,Fermous_LasAblPlume_2012, Bennaceur_SSExpKappaElectrons_2015, Murakami_MagFieldLasImp_2020}, implosion or expansion models in fusion contexts~\cite{Liberman_SSinZ-Pinch_1986,PégouriéRev_PelletInj_2007,Cassibry_SSConvergShock_2009,Arnold_SSPelletExp_2021}, dusty plasmas~\cite{Pillay_DustyPLasSSExp_1997,Shahmansouri_DustyPlasmaSS_2017}, expansion phenomena in aerospace and space plasmas~\cite{Low_SSMHD_1982,Arefiev_2AppsOfCollessExp_2009, Hu_SSbeamExpSpace-Aerospace_2020}, and ion acceleration schemes~\cite{MoraSelfSimilar2003,Naveen_MonoenIonSS_2008}. The hot electrons are often modeled by assuming a polytropic law for the electron temperature $T_e(n_e)$ as a closure relation. While polytropic closure models reproduce general features of the expansion observed in simulations and experiments in many contexts~\cite{Chan_ExpSSExp_1983,Wright_Stone_Samir_ExpInLabPlasma_1985,Denavit_ExpSimSScomp_1979}, they preclude self-consistent incorporation of external heating or cooling mechanisms into the plasma dynamics.

In Sec.~\ref{sec:setup} we demonstrate a range of different ``Self-similar closures" for the electron temperature that are more general than the polytropic description. We specialize to a one-dimensional planar geometry, which can be used to study the early-time intermediate asymptotic behavior for plasmas in arbitrary geometry when the length scale of the self similar-region $L(t)$ is much smaller than the dimensions of the target, i.e., for $L_0 \ll L(t) \ll R$, where $R$ is a characteristic length scale of the target. We discuss the permissible asymptotic temporal variation of $T_e$ controlled by a third free parameter, and derive a new three parameter self-similar system for the two-fluid $+$ Poisson equations. The obtained family is compared to previous self-similar solutions in the literature with charge separation, as well as those obtained under the quasineutrality assumption. A reduced system of equations with uniform electron temperature profiles relevant for laser heated plasma expansion is presented.

In sections~\ref{sec:LimitingSols} - \ref{sec:highEta}, detailed analysis is carried out for the special case of limiting self-similarity ($m\rightarrow0$) with uniform electron temperature $T_e(t)$. The solutions consistently incorporate the electron sheath effects and resolve the electrostatic field structures in the self-similar domain. The charge separation effects give rise to vastly different regimes of plasma dynamics from Coulomb explosion to quasineutral expansion, and resolve local features including density and velocity profile modulations of the expanding ion fluid and hydrodynamic shock-like structures close to the leading edge of the expanding ions. The solutions also elucidate the natural length scales of variation of the participating electron and ion fluids, and regimes of ion flow with subsonic and supersonic exit in the parameter space. The parametric variation of the solutions relates $L(t)$ and the characteristic rate of variation of the length scales, $\gamma$, to the different regimes of plasma dynamics, demonstrating continuous transition among them. The presented model thus brings together these various dynamical regimes and physical features into a single self-consistent framework. 

Section~\ref{sec:energetics} discusses how the energy deposited into the electrons by an external heating source is partitioned among ion kinetic energy, electrostatic field energy and electron thermal energy. The analysis is relevant to laser-plasma applications, such as determining efficiency in ion acceleration schemes. Lastly, in Sec.~\ref{sec:application}, we use the obtained solutions to analyze plasma dynamics at a target surface during high intensity laser-target interactions.

\section{Fluid Equations and Ansatz}\label{sec:setup}
We consider the planar expansion of an initially neutral, nonrelativistic, semi-infinite plasma slab occupying the negative half space $x<0$ and ionized to a charge state $Z$. The dynamics of the plasma of a length scale $L(t)$ is considered, and the electron temperature is assumed to have a spatiotemporal variation $T_e(x,t)$ in this expanding plasma region. $T_e(x,t)$ can be obtained from an energy equation for the electrons, that might include an external heating or cooling mechanism, or from an appropriate equation of state describing the physics.

In typical laser heated plasmas, the collisional mean free path of the electrons $\lambda_{\rm mf}\sim T_e^2n_e^{-1}$ increases faster than the expanding plasma length scale $L$. In such cases, the plasma eventually reaches a stage when the mean free path exceeds the plasma length scale, and the plasma becomes collisionless,
\begin{equation}\label{colLessApprox}
    L_0 < \lambda_{\rm mf0} = \sqrt{\frac{T_{e0}}{m_e}} \frac{1}{\nu_{ei0}}
\end{equation}
Here the subscript $0$ denotes the characteristic values in the collisionless stage. $\nu_{ei0}$ and $\lambda_{\rm mf0}$ are the characteristic (electron-ion) collision frequency and collisional mean free path in this stage. The mean free path can also be expressed in terms of the characteristic electron temperature $T_{e0}$ and density $n_{e0}$ as
\begin{equation}
    \lambda_{\rm mf0} \approx \frac{14}{Z \ln{\Lambda}} \left(\frac{T_{e0}}{100 ~\rm eV}\right)^2 \left(\frac{n_{e0}}{10^{20}~\rm {cm}^{-3}}\right)^{-1} \rm \mu m
\end{equation}
where $\ln{\Lambda}$ is the Coulomb logarithm. We describe the plasma evolution in this stage by the non-relativistic, collisionless hydrodynamic equations for the electrons and ions. The ion fluid is assumed to be cold and is coupled to the electron fluid through the electrostatic field determined by Poisson's equation. The coupled two-fluid system is governed by the equations:
\begin{subequations} \label{fluidlab:sys}
        \begin{align}
        & \frac{\partial n_\alpha}{\partial t} + \frac{\partial}{\partial x}(n_\alpha v_\alpha) = 0 \label{fluidlab:cont} \\
        & \frac{\partial v_i}{\partial t} + v_i\frac{\partial v_i}{\partial x} - \frac{Z e}{m_i}E = 0 \label{fluidlab:imom} \\
        & \frac{\partial v_e}{\partial t} + v_e\frac{\partial v_e}{\partial x} + \frac{e}{m_e}E + \frac{1}{m_en_e}\frac{\partial}{\partial x}(n_eT_e) = 0 \label{fluidlab:emom} \\
        & \frac{\partial E}{\partial x} = \frac{e}{\epsilon_0}(Zn_i-n_e)  \label{fluidlab:Poisson}
    \end{align}
\end{subequations}
where the subscript $\alpha = e,i$ denotes the quantities for the electron and ion fluids respectively. The first line represents two continuity equations for the ion and electron fluids, while the second and third equations are the fluid momentum equations for the ions and electrons respectively. The fourth equation is the Poisson equation, which gives the electrostatic field driving the expansion. The equation describing the electron fluid temperature $T_e(x,t)$ closes these hydrodynamic equations. The type of closure depends on the method for solving the system.

To obtain universal behavior of the expansion especially at later times, we choose a self-similar Ansatz. Self-similar solutions provide the behavior when a characteristic time-dependent length scale $X(t)$ of the dynamics much exceeds its initial value,
\begin{equation}\label{simCond}
    X(t)\gg X_0=X(t=0) ,
\end{equation}
thus representing the ``intermediate asymptotic'' dynamics of the system at sufficiently large times when condition~\eqref{simCond} holds. The generic length scale $X(t)$ can take the value of a natural length scale in the system, a length scale introduced from the boundary conditions, or a combination of these with dimension of length. In this limit, with $\sfrac{X_0}{X(t)}$ manifestly small, the dynamics becomes insensitive to the details of the initial plasma profiles. Such solutions are thus seen as universal attractors for a wide class of initial and boundary conditions for which the system asymptotically depends on the single time dependent length scale $X(t)$. For such systems the dynamics of all degrees of freedom depend on $X(t)$, and the system develops a scale invariance with respect to the time-dependent scaling coordinate
\begin{equation} \label{SSvargen}
    \xi=\frac{x}{X(t)},
\end{equation}
where $x$ is the dimension of expansion.

If the dynamics asymptotes towards a non-zero finite limit independent of $X_0$, then the quantities of the system in this limit can be expressed as univariate functions of $\xi$, and the system is said to exhibit a complete self-similarity. Complete self-similar solutions require that no intrinsic physical length scales survive in the asymptotic stage other than the global scale $X(t)$. If no such non-trivial, finite limit exists, the system exhibits an incomplete self-similarity in which, the intermediate asymptotic behavior of the physical quantities $q_j$ scale with some exponent of the parameter $\sfrac{X_0}{X(t)}$, i.e., $q_j \sim \left(\sfrac{X_0}{X(t)}\right)^{m_j}$. For every choice of the scaling exponent(s) $m_j$ leading to self-similar solutions of the governing equations, the equations might lead to a distinct form of temporal evolution of $X(t)$. For systems whose intermediate asymptotics are governed by a particular choice of the exponents, any length scale must asymptotically evolve with the temporal form of the corresponding $X(t)$. The incomplete self-similar solutions represent a continuous family of universality classes parametrized by $m_j$ and $X_0$. Each member of the family corresponds to the intermediate asymptotic evolution of all systems with an initial scale $X_0$ and the same asymptotic time-dependent scaling properties~\cite{Barenblatt_1996}.

In the introduction, we identified three characteristic length scales for heated plasma expansion, $\lambda_D(t)$, $\lambda_s(t)$ and $L(t)$, and the characteristic length $X(t)$ can be any combination of the three. Consistent self-similar scaling solutions are possible if and only if these three lengths are kept strictly proportional,
\begin{equation} \label{scalesEvol}
    \frac{\lambda_s(t)}{\lambda_{s0}} = \frac{\lambda_D(t)}{\lambda_{D0}} = \frac{L(t)}{L_0}~.
\end{equation}
The two equalities imply two free parameters, e.g. $\sfrac{\lambda_{D0}}{L_0}$ and $\sfrac{\lambda_{s0}}{L_0}$, that are invariant during the self-similar evolution. Their values are set by the plasma state at a reference time ($t=0$ hereafter) in the late time collisionless stage.

Non-zero $\lambda_D\sim T_e^{\frac{1}{2}}n_e^{-\frac{1}{2}}$ as an intrinsic length scale prohibits complete self-similar solutions for the system, because Eq.~\eqref{scalesEvol} implies
\begin{equation} \label{TeNeScale}
    T_e^{\frac{1}{2}}n_e^{-\frac{1}{2}} \sim X
\end{equation}
The electron density and temperature must retain explicit time dependent scalings with respect to $X(t)$ such that condition~\eqref{TeNeScale} is obeyed. We find incomplete self-similar solutions by including power-law dependence on $(\sfrac{X_0}{X})$ in the density, temperature and electric field through an Ansatz of the form,
\begin{subequations} \label{SSanz:denVel}
       \begin{align}
        n_\alpha &= n_{\alpha0} \left(\frac{X_0}{X}\right)^m N_\alpha(\xi) ,\\
        v_\alpha &= \dot{X} \mkern4mu V_\alpha(\xi) ,\\
        T_e &= \frac{m_i X_0^2 \gamma^2}{Z} \left(\frac{X_0}{X}\right)^{m-2} \mkern4mu \Theta(\xi) \label{genanz:temp}\\
        eE &= \frac{m_i X_0 \gamma^2}{Z} \left(\frac{X_0}{X}\right)^{m-1}\mkern4mu \mathcal{E}(\xi)
    \end{align}
\end{subequations}
where $n_{e0} = Z n_{i0}$ is the characteristic electron density of the plasma, $\gamma = \frac{\dot{X}}{X_0}\big|_{t=0}$ is the characteristic rate of expansion, and $Z$ is the ion charge state. The fluid velocities are normalized to the expansion speed of the length scale $X$. The electron thermal energy and work done by the electric field on the ions are written in units of the characteristic ion inertial energy, $m_i \dot{X}^2|_{t=0}$, associated with the expansion speed $\dot{X}|_{t=0}$. The parameter $m$ is subject to constraints emerging from the choice of physical solutions, such as mass or energy flux at a boundary during this stage, which will be discussed and chosen a bit later. Consistency of these time evolutions for the quantities with the self-similarity constraint Eq.~\eqref{scalesEvol} results in a range of possible closures for $T_e$ at this stage.

Substituting the ansatz into Eqs.~\eqref{fluidlab:sys} leads to the equation for $X(t)$,
\begin{equation} \label{XoftEvol}
    X \dfrac{d^2X}{dt^2} = \left(1-\frac{m}{2}\right) \left(\dfrac{dX}{dt}\right)^2 ,
\end{equation}
and the fluid equations reduce to a system of Ordinary Differential Equations(ODEs) in the self-similar coordinate $\xi$,
\begin{subequations} \label{fluidSS:gen}
        \begin{align}
        & -mN_\alpha+\left(V_\alpha-\xi\right)\frac{dN_\alpha}{d\xi} + N_\alpha\frac{dV_\alpha}{d\xi} = 0 \label{fluidSS:cont} \\
        & \left(1-\frac{m}{2}\right)V_i + \left(V_i-\xi\right)\frac{dV_i}{d\xi} - \mathcal{E} = 0 \label{fluidSS:imom} \\
        & \mu\left[\!\left(\!1-\frac{m}{2}\right)\!V_e + \left(V_e-\xi\right)\!\frac{dV_e}{d\xi}\!\right] + \frac{1}{N_e}\frac{d}{d\xi}(N_e\Theta) + \mathcal{E} = 0  \\
        & \frac{d\mathcal{E}}{d\xi} = \eta (N_i-N_e)
    \end{align} 
\end{subequations}
Here $\mu=\sfrac{Zm_e}{m_i}$ is the electron-to-ion mass ratio and $\eta = \left(\sfrac{\omega_{pi0}}{\gamma}\right)^2$, with $\omega_{pi0}=\left(\tfrac{n_{i0}Z^2e^2}{m_i\epsilon_0}\right)^{\sfrac{1}{2}}$ the characteristic ion plasma frequency in the plasma. The parameter $\eta$ measures the expansion timescale relative to the characteristic ion response time $\omega_{pi0}^{-1}$, and remains one of two primary parameters determining the expansion physics as mapped by Fig.~\ref{fig:regSchematic}.

The solutions of Eq.~\eqref{XoftEvol} determine the asymptotic evolution of the scale $X(t)$ that admit self-similar solutions for the electrostatically coupled two-fluid system. For $X(t=0) = X_0 \neq 0$, the resulting evolution is
\begin{equation} \label{genXoft}
    X(t) = X_0
    \begin{cases}
        \left(1+\frac{m}{2}\gamma t\right)^{\frac{2}{m}}& m \neq 0 \\
        \exp{\left(\gamma t\right)}& m = 0 \\
    \end{cases}
    \end{equation}
The spatial structure of the self-similar solutions is governed by the reduced ODE system~\eqref{fluidSS:gen}. These equations possess singular points at which $V_\alpha=\xi$. At such a point $\xi_f$, Eqs.~\eqref{fluidSS:cont} and \eqref{fluidSS:imom} imply
\begin{subequations} \label{XifValsGen}
        \begin{align}
         V_i(\xi_f) &= V_{f} = \xi_f  \label{viXifValsGen}\\
         \mathcal{E}(\xi_f) &= \mathcal{E}_f = \left(1-\frac{m}{2}\right)\xi_f
    \end{align} 
\end{subequations}
Condition~\eqref{viXifValsGen} implies that the local ion fluid velocity equals the velocity of the location $x_f(t) = X(t) \xi_f$, i.e., $v_i\left(x_f(t)\right) = \dot{x}_f$. Thus, $x_f(t)$ represents a moving boundary of the self-similar solutions across which there is no ion mass flux.

The class of admissible electron temperature evolutions compatible with the self-similar framework follow from Eqs.~\eqref{genanz:temp} and \eqref{genXoft}. These equations can be used to express the electron temperature profile in the form
\begin{equation}\label{genTvar}
    T_e = T_{e0}\frac{\Theta(\xi)}{\Theta_0}\left(\frac{X(t)}{X_0}\right)^{2\left(1-\frac{m}{2}\right)}
    \end{equation}
where 
\begin{equation}
\Theta_0=\left(\tfrac{C_{s0}}{\gamma X_0}\right)^2    ,
\end{equation}
and $C_{s0}=\left(\frac{ZT_{e0}}{m_i}\right)^{\sfrac{1}{2}}$ is the characteristic speed of sound at $t=0$ with the reference temperature $T_{e0}$ chosen at a self-similar boundary $\xi=\xi_c$, i.e. $T_{e0}=T_e\left(\xi_c,t=0\right)$. The spatial profile $\Theta(\xi)$ is determined by the relevant heating/cooling mechanism for the plasma (and/or an appropriate equation of state) and, once specified, closes the self-similar system~\eqref{fluidSS:gen}. We note that Eqs.~\eqref{fluidSS:gen} are invariant under the mapping
\begin{equation} \label{scalingMap}
            \xi \rightarrow a\xi, \quad V_\alpha \rightarrow aV_\alpha, \quad \mathcal{E} \rightarrow a\mathcal{E}, \quad \Theta \rightarrow a^2\Theta 
\end{equation}
If the closure for $\Theta(\xi)$ respects the same scale invariance, the system admits a further reduction. This will be demonstrated for the special case of uniform heating in the Sec.~\ref{sec:uniHeating}.

The self-similar ansatz along with the scale evolution Eq.~\eqref{genXoft} represent a form of $t-\xi$ separation that provide self similar solutions of system~\eqref{fluidlab:sys} ensuring that Eq.~\eqref{scalesEvol} is satisfied. Evaluating $\lambda_D$ and $\lambda_s$ at the self-similar boundary $\xi_c$ (with $|\xi_c|=\sfrac{L(t)}{X(t)}$ and $N_e(\xi_c)=N_{e0}$) gives
\begin{subequations}
\begin{align}
    \lambda_{\!D}(\xi_c,t) &= \left(\mkern-3mu\frac{\epsilon_0 T_e(\xi_c,t)}{n_e(\xi_c,t) e^2}\mkern-3mu\right)^{\mkern-5mu \frac{1}{2}} \mkern-6mu = \frac{\lambda_{D0}}{\mkern-2mu\sqrt{N_{e0}}} \frac{X(t)}{X_0} \\
    \lambda_s(\xi_c,t) &= \left(\mkern-3mu\frac{Z T_e(\xi_c,t)}{m_i}\mkern-3mu\right)^{\mkern-5mu\frac{1}{2}}\mkern-4mu\frac{L}{\dot{L}} = \lambda_{s0} \frac{X(t)}{X_0} \label{lambdaS}
\end{align}
    \end{subequations}
with $\lambda_{D0} = \left(\frac{\epsilon_0 T_{e0}}{n_{e0}e^2}\right)^{\sfrac{1}{2}}$ and $\lambda_{s0}= \left(\frac{Z T_{e0}}{m_{i} \gamma^2}\right)^{\sfrac{1}{2}}$. These relations show that the self-similar ansatz~\eqref{SSanz:denVel} ensures that $\lambda_D$, $\lambda_s$ and $L$ scale in proportion to $X(t)$. Systems which asymptotically approach one of these self-similar solutions must obey the temporal law~\eqref{genXoft}  for $L(t)$ and the temperature scaling~\eqref{genTvar} for the corresponding $m$. 

The above construction yields a three parameter family of self similar solutions, whose qualitative behavior is governed by the boundary conditions and the choice of electron temperature variation. The two independent free parameters $\sfrac{\lambda_D}{L}$ and $\sfrac{\lambda_s}{L}$ resulting from constraint~\eqref{scalesEvol} are encoded by the choices of $|\xi_c|$, $\eta$ and $\Theta_0$. Since $X(t)$ is a generic length scale that can be formed from $\lambda_D$, $\lambda_s$ and $L$ (without loss of generality), only two independent dimensionless parameters arising from these quantities enter the self-similar dynamics. A third independent parameter is the exponent $m$, which governs the asymptotic temporal variation of $T_e$ and $L$. In particular, $m=2$ corresponds to a constant temperature profile and a linear temporal growth of $L(t)$, while $m<2$ ($>2$) yields an increasing (decreasing) $T_e$ in time. Apart from these temporal evolutions, $m$ also controls the mass flux into the self-similar domain. The total mass $\mathcal{N}_\alpha$ of the species $\alpha$ in the self-similar region is given by,
\begin{equation}
    \mathcal{N}_\alpha = \mathcal{N}_{\alpha 0} \left(\frac{X(t)}{X_0}\right)^{1-m}
\end{equation}
where $\mathcal{N}_{\alpha 0}$ is the initial mass per unit area in the self-similar region ranging from $\xi_c$ to $\xi_f$,
\begin{equation}
    \mathcal{N}_{\alpha 0} = n_{\alpha0} X_0 \int_{\xi_c}^{\xi_f} N_\alpha(\xi) d\xi
\end{equation}

\subsection{Uniform Temperature} \label{sec:uniHeating}
In typical laser heated plasmas, the electron temperature in the hot expanding region is nearly uniform due to the fast heat conduction at high electron temperatures. Electron heat flux in the hot collisionless plasma is similar to but strictly smaller than the saturating (or free streaming) heat flux $q_0 = \frac{3}{2}n_e T_e v_e$, where $v_e = \sqrt{\sfrac{T_e}{m_e}}$ is the electron thermal velocity. The free-streaming bound corresponds to electron energy being primarily transported by their ballistic motion in the characteristic electron crossing time $\tau_{\rm{free}} = \sfrac{L}{v_e}$. If this timescale is much smaller than the timescale of expansion $\sfrac{1}{\gamma}$, then the bulk electron population in the expanding region gets kinetically mixed and develops an almost uniform temperature $T_e(t)$. Thus, in the collisionless self-similar stage of a heated plasma, the electron temperature may be approximated by a spatially uniform temperature when the following criterion is met:
\begin{equation} \label{condApprox}
    \gamma \tau_{\rm free} \ll 1
\end{equation}

The governing equations for the self-similar plasma dynamics with spatially uniform electron temperature profiles $T_e \mkern-6mu = \mkern-4mu T_e(t)$ can be obtained by setting $\Theta(\xi) \mkern-6mu = \mkern-4mu \Theta_0$ in Eqs.~\eqref{SSanz:denVel} - \eqref{genTvar}. Equation~\eqref{genTvar} simplifies the permissible electron temperature profiles to
\begin{equation}\label{genTvar_uni}
    T_e = T_{e0}
    \begin{cases}
        \left(1+\frac{m}{2}\gamma t\right)^{2\left(\frac{2}{m}-1\right)}& m \neq 0 \\
        \exp{\left(2\gamma t\right)}& m = 0 \\
    \end{cases}
    \end{equation}
    
Since the system is invariant under mapping~\eqref{scalingMap}, we introduce the scale-invariant quantities
\begin{subequations} \label{scaledVars}
        \begin{align}
         \zeta &= \frac{\xi}{ \sqrt{\Theta_0}} \\
         P_{\alpha} &= \frac{V_{\alpha}}{ \sqrt{\Theta_0}} \\
         Q &= \frac{\mathcal{E}}{\sqrt{\Theta_0}}
    \end{align} 
\end{subequations}
to replace $\xi$, $V_{\alpha}$ and $\mathcal{E}$ in the system. Furthermore, for typical laser heating contexts, the heating timescale is much larger than electron inertial response time,
\begin{equation} \label{ESapprox}
    \omega_{pe0}\gg\gamma
\end{equation}
In such cases, the electron inertia can be neglected ($\mu\approx0$), and the electron density profile is governed by the electrostatic field given by the Poisson equation. The electron continuity equation then provides the electron velocity profile. Setting $\mu = 0$ in Eqs.~\eqref{fluidSS:gen} and dropping the electron continuity equation, the governing equations reduce to
\begin{subequations} \label{SSuniGen:sys}
        \begin{align}
        & -m N_i + \left(P_i-\zeta\right)\frac{dN_i}{d\zeta} + N_i\frac{dP_i}{d\zeta} = 0 \label{SSuniGen:cont} \\
        & \left(1-\frac{m}{2}\right)P_i + \left(P_i-\zeta\right)\frac{dP_i}{d\zeta} - Q = 0 \label{SSuniGen:imom} \\
        & \frac{1}{N_e}\frac{dN_e}{d\zeta} + Q = 0 \label{SSuniGen:emom} \\
        & \frac{dQ}{d\zeta} = \eta (N_i-N_e) \label{SSuniGen:Poisson}
    \end{align}
\end{subequations}

In the transformed system~\eqref{SSuniGen:sys}, the new self-similar variable is
\begin{equation} \label{scaledSSvarGen}
            \zeta  = \frac{x}{ \sfrac{C_{s0}}{\gamma}} \frac{X_0}{X(t)}
            =            \frac{x}{\lambda_s(t)}
\end{equation}
Comparing Eqs.~\eqref{SSvargen} and \eqref{scaledSSvarGen}, the transformation \eqref{scaledVars} is equivalent to setting $X(t)=\lambda_s(t)$ where,
\begin{align} \label{lambdaS}
            \lambda_s(t)
            &= \lambda_{s0}
            \begin{cases}
            \left(1+\frac{m}{2}\gamma t\right)^{\sfrac{2}{m}}& m \neq 0 \\
        \exp{\left(\gamma t\right)}& m = 0
            \end{cases} \\
            \mkern8mu &= \lambda_{s0}+\int_0^t C_s(t') dt'
\end{align}
with $\lambda_{s0} = \sfrac{C_{s0}}{\gamma}$. With \( C_{s}(t) \) as the ion sound speed in the plasma, $\lambda_s(t)$ represents the typical distance traversed by ion-acoustic waves in the plasma in time $t$. Lastly, apart from serving as a measure for the ion response rate in the expansion timescale, the parameter $\eta$ in Eqs.~\eqref{SSuniGen:sys} also quantifies the ion-acoustic correlation length $\lambda_s$ in terms of the Debye length $\lambda_D$. Using $\lambda_{s0}=\frac{C_{s0}}{\gamma}$ and $\lambda_{D0}= \left(\frac{\epsilon_0 T_{e0}}{n_{e0} e^2}\right)^{\sfrac{1}{2}}$, we may write
\begin{equation} \label{eta}
            \eta = \left(\!\frac{\omega_{pi0}}{\gamma}\!\right)^{\!2} = \left(\!\frac{\sfrac{C_{s0}}{\gamma}}{\lambda_{D0}}\!\right)^{\!2} = \left(\!\frac{\lambda_{s0}}{\lambda_{D0}}\!\right)^{\!2} = \left(\!\frac{\lambda_s(t)}{\lambda_D(t)}\!\right)^{\!2}
\end{equation}
Thus $\sfrac{1}{\sqrt{\eta}}$ gives the variation scale of space charge separation, $\lambda_D(t)$, relative to $\lambda_s(t)$.

\subsection{Relationship to previous work}
For quasineutral plasma expansion, self-similar solutions of both first and second kind with a polytropic closure have been studied extensively in the literature~\cite{SACK1987311,GurevichSelfSimilar,mora1979self,Schmalz_SS_type2,Baitin_SSExpRelativistic_1998,Huang_QNSSGenIsoTe_2008}. Such solutions are obtained in the limit of a vanishing Debye length $\sfrac{\lambda_D}{L} \rightarrow 0$ which effectively reduces the system~\eqref{fluidlab:sys} to that of ideal gas hydrodynamics, thus relaxing constraint~\eqref{TeNeScale}. The predicted expansion dynamics is valid when the electron temperature does not vary much faster than the typical response time of the ions, and the plasma expands much faster than the Debye length. The quasineutrality assumption obscures features in the plasma profiles resulting due to electron-ion charge separation, such as the structure of collisionless shocks near the expanding ion front~\cite{DispersiveHydroShocksgurevich1984,Denavit_ExpSimSScomp_1979,Mora_SScollessExp_2005}. By retaining a finite Debye length in the solutions, our framework models scenarios where the electron temperature varies on a timescale comparable to that of expansion ($\gamma$), and resolves such charge separation induced features in the spatial variations of the plasma profiles.

An alternate route to simplifying Eqs.~\eqref{fluidlab:sys} with Ansatz \eqref{SSanz:denVel} is to find solutions with $V_\alpha=\xi$, which can be shown to necessitate $m$ to take a value of $1$ for the planar self-similar solutions of Eqs.~\eqref{fluidlab:sys}. In general, for one-dimensional solutions with $V_\alpha=\xi$ in arbitrary geometry, $m=\nu$, where $\nu=\{1,2,3\}$ is the number of spatial dimensions in the expansion geometry. These are incomplete self-similar solutions that correspond to scenarios where there is no mass flux into the expanding self-similar region. Such solutions were analyzed by Murakami and Basko~\cite{MurakamiSelfSimilar2006} for the expansion of a finite-sized plasma with a spatially uniform electron temperature $T_e=T_e(t)$. In this treatment, the evolution of $T_e(t)$ was assumed to be governed by a polytropic law of the form $T_e n_e^{1-\Gamma}|_{\xi=0} = T_{e0} n_{e0}^{1-\Gamma}$, where $\Gamma$ is the polytropic index. Under the constraints of mass conservation and choice of the polytropic form of evolution for $T_e$, the self-similar solutions were realized to exist only for $\Gamma=2-\sfrac{2}{\nu}$.

\section{Solutions for an expanding Plasma Slab} \label{sec:LimitingSols}
To demonstrate the dynamics predicted by our self-similar framework, in the following sections of this article we analyze the late stage heated expansion of an initially neutral plasma slab occupying the negative half-space $x<0$. We assume that the dynamics comprises a perturbed plasma region near the initial plasma-vacuum interface $x=0$, and the ion fluid beyond this region (in the negative-$x$ direction) is at the unperturbed density $n_{i0}$. The boundary between the two regions $x_c(t)$ propagates towards the unperturbed plasma, i.e. in the negative-$x$ direction. We choose $L(t)$ to denote the distance of this boundary from $x=0$ in the late expansion stage, $x_c(t)=-L(t)$, and $t=0$ to denote the onset of this stage. $L(t)$ has an initial value of $L(t=0) = L_0$ due to some initial stage dynamics at $t<0$ not included in our model. For laser-matter interactions, this initial dynamics includes ionization and initial heating as described in the introduction, and produces a temperature $T_{e0}$ and the scales $L_0$, $\lambda_{D0}$ and $\lambda_{s0}$ at $t=0$.

Matching the ion density in the expanding region to its unperturbed value at $x_c(t)$, $n_i(x_c)=n_{i0}$ requires that the ion density has no explicit time dependence. This makes the limiting self-similar solution ($m\rightarrow0$) a suitable choice for modeling the late stage evolution of the expanding region. Using Eqs.~\eqref{SSanz:denVel}, the self-similar Ansatz in this case can be written as
\begin{subequations} \label{SSanz:exp}
       \begin{align}
        n_\alpha &= n_{\alpha_0}N_\alpha(\zeta) \label{SSanzExp:n} \\
        v_\alpha &= C_{s0} \exp{\left(\gamma t\right)} P_\alpha(\zeta) \\
        T_e &=  T_{e0}\exp{\left(2\gamma t\right)} \label{SSanzExp:Te} \\
        eE &= T_{e0}\frac{1}{\lambda_{s0}} \exp{\left(\gamma t\right)} Q(\zeta)
    \end{align}
\end{subequations}
and, Eqs.~\eqref{SSuniGen:sys} give
\begin{subequations} \label{SSuni:sys}
        \begin{align}
        & \left(P_i-\zeta\right)\frac{dN_i}{d\zeta} + N_i\frac{dP_i}{d\zeta} = 0 \label{SSuni:cont} \\
        & P_i + \left(P_i-\zeta\right)\frac{dP_i}{d\zeta} - Q = 0 \label{SSuni:imom} \\
        & \frac{1}{N_e}\frac{dN_e}{d\zeta} + Q = 0 \label{SSuni:emom} \\
        & \frac{dQ}{d\zeta} = \eta (N_i-N_e) \label{SSuni:Poisson}
    \end{align} 
\end{subequations}
in the self-similar variable $\zeta = \frac{x}{\lambda_{s0}\exp\left(\gamma t\right)}$.

The limiting self-similar equations describe the asymptotic dynamics in a late stage when a heating mechanism produces a $T_e$ variation of the form Eq.~\eqref{SSanzExp:Te} and the perturbed plasma region expands exponentially, $L(t)=L_0 \exp{(\gamma t)}$. The exponential forms of explicit time dependence for the length scales and the quantities in Eq.~\eqref{SSanz:exp} are a result of the strict proportionality of length scales necessitated by Eq.~\eqref{scalesEvol}. However, in practice, these solutions also approximate the late stage dynamics when $\lambda_D(t)$ and $\lambda_s(t)$ do not grow much faster or slower than $L(t)$. For such systems Eq.~\eqref{scalesEvol} approximately hold in this stage.

The ion fluid boundary denoted by Eqs.~\eqref{XifValsGen} represents the expanding ion front until which the ion fluid extends. At this boundary $\zeta_f$,
\begin{equation} \label{ZetafVals}
         P_{if} = Q_f =  \zeta_f ~ ,
\end{equation}
where $P_{if} = P_i\left(\zeta_f\right)$ and $Q_f = Q\left(\zeta_f\right)$. The region $\zeta>\zeta_f$ is occupied by an electron sheath governed by the electron momentum and Poisson equations. Setting $N_i = 0$ in Eqs.~\eqref{SSuniGen:emom} and \eqref{SSuniGen:Poisson} we obtain,
\begin{subequations} \label{SSuni:sheath}
        \begin{align}
        & \frac{1}{N_{e\rm sh}}\frac{dN_{e\rm sh}}{d\zeta} + Q_{\rm sh} = 0  \\
        & \frac{dQ_{\rm sh}}{d\zeta} = -\eta N_{e\rm sh}
    \end{align} 
\end{subequations}
where, $N_{e\rm sh}$ and $Q_{\rm sh}$ are the electron density and the electric field in the sheath respectively. Continuity of these quantities at $\zeta_f$ requires
\begin{equation} \label{zetaf-cont}
       N_{e\rm sh}(\zeta_f) = N_{e}(\zeta_f) = N_{ef} \quad ; \quad Q_{\rm sh}(\zeta_f) =  Q_f
\end{equation}
Using $\lim_{Q_{\rm sh}\to0}N_{e\rm sh}\to0$, Eqs.~\eqref{SSuni:sheath} and \eqref{zetaf-cont} can be used to obtain the electron sheath profiles,
\begin{subequations} \label{sheath-zeta}
    \begin{align}
         Q_{\rm sh}(\zeta) &= Q_f\left[\left(\zeta-\zeta_f\right)\frac{Q_f}{2}+1\right]^{-1} \\
         N_{e\rm sh}(\zeta) &= N_{ef}\left[\left(\zeta-\zeta_f\right)\frac{Q_f}{2}+1\right]^{-2} \label{sheath-Nes}
    \end{align}
\end{subequations}
and the matching condition at $\zeta_f$,
\begin{equation} \label{match}
      N_{ef}=\frac{Q_f^2}{2\eta}
\end{equation}

In the expanding ion fluid region $\zeta_c\leq\zeta<\zeta_f$, the solutions are obtained by matching the ion fluid properties and the electric field to their corresponding values in the unperturbed plasma region at $x_c(t)$. At the boundary $x_c(t)$, corresponding to the self-similar coordinate
\begin{equation} \label{zeta_c-exp}
        \zeta_c = -\frac{L(t)}{\lambda_s(t)} = -\frac{L_0 \gamma}{C_{s0}} ,
\end{equation}
the ion and electron densities, ion velocity and electric field take the values
\begin{equation} \label{coreBCs}
       N_i \Big|_{\zeta_c}\! = 1 \mkern12mu;\mkern12mu  N_e \Big|_{\zeta_c}\! = N_{e0} \mkern12mu ;\mkern12mu  P_i \Big|_{\zeta_c}\! = 0  \mkern12mu;  \mkern12mu Q \Big|_{\zeta_c} \! = 0
\end{equation}
The electron density at this boundary $N_{e0}$ assumes a value $0<N_{e0}<1$ due to some electron motion in the unperturbed region. This electron motion near $\zeta_c$ is unresolved by our model since the electron's inertial response has been neglected. With the electrons in electrostatic equilibrium with the field, the sheath field is related to the value of the electron density at the ion expansion front through Eq.~\eqref{match}. Thus, condition~\eqref{match} results in unique determination of the boundary value $N_{e0}$ at $\zeta_c$.

Conditions~\eqref{colLessApprox}, \eqref{condApprox} and \eqref{ESapprox} place the following constraints on the parameters in the model for its validity,
\begin{subequations}
        \begin{align}
            \eta &\gg \frac{Z m_e}{m_i N_{e0}} = \frac{\mu}{N_{e0}}\\
            |\zeta_c| &\ll \min\left\{\frac{\lmfp^0}{\lambda_s} , \frac{1}{\sqrt{\mu}}\right\} = \frac{1}{\sqrt{\mu}}\min\left\{\frac{\gamma}{\nu_{ei}} , 1\right\}
        \end{align}
\end{subequations}
where Eq.~\eqref{zeta_c-exp} and $L(t) = L_0\exp{\left(\gamma t\right)}$ have been used. With $\nu_{ei}\propto T_e^{-\sfrac{3}{2}}$, the range of validity of the model steadily increases with time as $T_e$ increases exponentially.

\subsection{Qualitative view of parametric regimes} \label{sec:Qual}
\begin{figure*}[htb]
    \centering
    \includegraphics[width=1\linewidth]{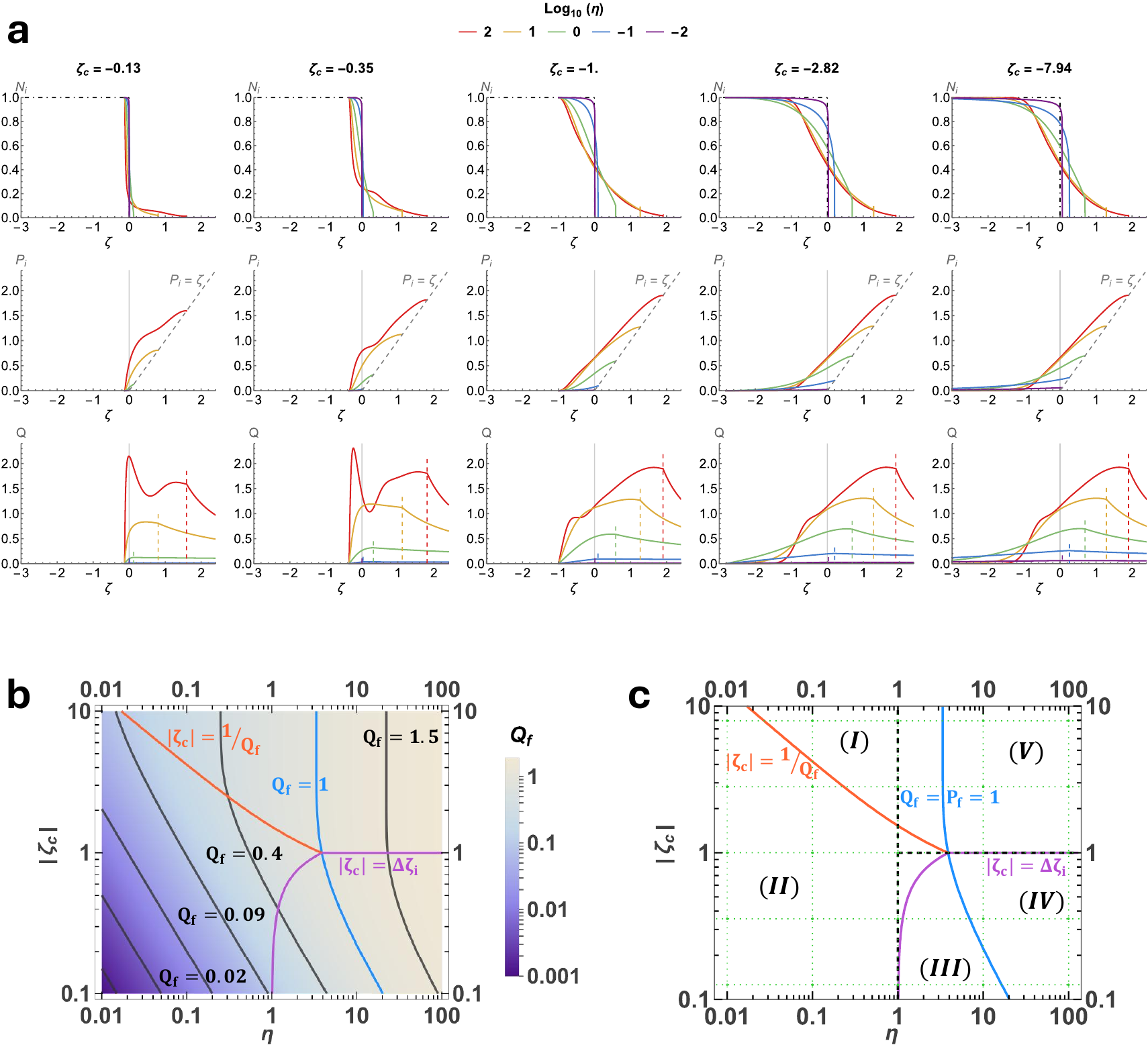}
    \caption{(a) Profiles for the ion density $N_i$, ion velocity $P_i$ and electrostatic field $Q$ for $\eta \in \left\{10^{-2},10^{-1},1,10,10^{2}\right\}$ plotted in different colors, for $-\zeta_c \in \left\{0.13,0.35,1,2.82,7.94\right\}$ corresponding to each column from the left. The dashed gray lines in the $P_i$ plots represent $P_i=\zeta$ while the colored dashed lines in the $Q$ plots represent $\zeta=\zeta_f$ for the solutions of the corresponding colors. The $\zeta=0$ interface confining the initial plasma to the negative half-space is shown with the gray solid line in the $P_i$ and $Q$ plots, and the black dot-dashed lines in the $N_i$ plots is the initial density profile. $(b)$ The parametric variation of $Q_f$ with respect to $\eta \in [10^{-2},10^{2}]$ and $|\zeta_c| \in [0.1,10]$ with contours of fixed $Q_f$ shown in black and blue (for $Q_f=1$) lines. In $(b)$ and $(c)$, orange and purple contours are $|\zeta_c|=\zeta_{Df}$ and $|\zeta_c|=\Delta\zeta_i$. $(c)$ Schematic of the $\eta-|\zeta_c|$ space, showing the $5$ asymptotic regimes. The points in the parameter space for which the profiles are plotted in $(a)$ are marked with green dots, and the sonic ion exit contour $P_f = 1$ is shown in blue}
    \label{fig:paramVarPlots}
\end{figure*}

Equations~\eqref{SSuni:sys} can be solved along with boundary conditions~\eqref{coreBCs} and matching condition~\eqref{match}, to obtain the plasma dynamics for $\zeta_c\leq\zeta<\zeta_f$, for different values of the parameters $\eta$ and $\zeta_c$. These equations were numerically solved for $\eta$ and $\zeta_c$ in the ranges $\left[10^{-2},10^2\right]$ and $\left[0.1,10\right]$ respectively, and the profiles of $N_i$, $P_i$ and $Q$ for some of the solutions are shown in Fig.~\ref{fig:paramVarPlots}$(a)$. Each column corresponds to the value of $\zeta_c$ mentioned on the top, and each plot shows the profiles for $\eta \in \{10^{-2},10^{-1},1,10,10^{2}\}$. The dot-dashed line in the ion density plots represents the initially neutral plasma slab, and the gray vertical line in the ion velocity and electric field plots corresponds to the initial vacuum interface. The contour $P_i=\zeta$, on which $\zeta_f$ lies, is shown with a dashed line in the $P_i$ plots. The numerical integrations were carried out until $\zeta_f-\epsilon<\zeta_f$ to avoid numerical artifacts at $\zeta_f$ with the end point difference $\epsilon=10^{-8}|\zeta_c|$. The colored dashed lines in the plots for $Q$ correspond to the boundary $\zeta_f$ that separates the electron-ion plasma region from the electron sheath beyond it.

The self-similar scale $\lambda_s(t)$ and the time-dependent normalizations of the velocity and electrostatic field in Ansatz~\eqref{SSanz:exp} is fully determined by the electron temperature profile. Thus, the various plots in Fig.~\ref{fig:paramVarPlots}$(a)$ can be interpreted as snapshots of the velocity, electric field and normalized ion density $(\sfrac{n_i}{n_{i0}})$ profiles at a time $t$ with electrons at a temperature $T_e(t)$. The parameters $\eta \sim n_{i0}$ and $|\zeta_c| \sim L$ then describe varying plasma conditions corresponding to different unperturbed plasma densities and different lengths of the perturbed plasma domain. The parameter $\eta = \left(\sfrac{\omega_{pi0}}{\gamma}\right)^2$ determines the response rate of the expanding ion fluid in the heating timescale, while $|\zeta_c|$ characterizes the extent of the perturbed plasma region relative to the ion correlation length $\lambda_s(t)$. At lower plasma densities (producing lower $\eta$) the Debye length is large, and the ion and electron fluids are weakly coupled through the Poisson equation. This results in weak electrostatic fields in the ion plasma and electron sheath regions. As a consequence, the ions do not accelerate substantially, exhibiting a slow response on the timescale of electron heating. This behavior is observed for the $\eta \ll 1$ profiles in Fig.~\ref{fig:paramVarPlots}$(a)$, where the electric field is very weak, and the ion profiles have marginal deviations from the initial profile. With increasing $\eta$, the sheath field gets stronger and more localized. The stronger fields accelerate the ions to higher velocities, as evidenced by the ion velocity plots with increasing $\eta$. The localization of the field in the $\eta \gg 1$ limit results in space charge oscillations, producing modulations in the ion density and velocity profiles.

The distance $L$ of the boundary between the perturbed and unperturbed plasma determines the mass of the plasma participating in the dynamics. The electron mass per unit area in the expanding ion fluid and electron sheath region is given by $|\zeta_c|$, normalized to $n_{e0}\lambda_s(t)$. When $|\zeta_c| \ll 1$, the very small mass of electrons spread across these expanding regions creating a diffuse electron plasma of a density much lower than the unperturbed plasma density. This results in plasma profiles with a large charge separation in the ion fluid region near the unperturbed plasma boundary, which will be detailed further in sections~\ref{sec:lowEta} and \ref{sec:highEta}. In the high$-|\zeta_c|$ limit, the plasma near the unperturbed plasma boundary is farther than the characteristic distance over which the expansion wave propagates from the initial vacuum interface. The plots in Fig.~\ref{fig:paramVarPlots}$(a)$ illustrate that this results in the plasma profiles becoming insensitive to the value of $|\zeta_c|$ at large $|\zeta_c|$. These various qualitative aspects arising due to varying $\eta$ and $|\zeta_c|$ exemplify the effects that the relative magnitudes of the scales $L$, $\lambda_D$ and $\lambda_s$ have on the expansion dynamics. These relative magnitudes influence the characteristics of the driving electrostatic field, and produce dynamically distinct expansion regimes.

The above discussion demonstrates the crucial role played by the sheath field and the mass of the expanding plasma $|\zeta_c|$ in dictating the qualitative nature of the dynamics. An essential descriptor of the sheath field is its value at the ion expansion front, $Q_f$, that governs both its strength and its variation scale. Equations~\eqref{sheath-zeta} suggest that the electron density and the sheath field beyond $\zeta_f$ vary on a $\zeta-$scale of $\Delta\zeta_{\rm sh}=\sfrac{2}{Q_f}$. This scale indeed corresponds to the Debye length scale at $\zeta_f$, $\lambda_{Df}$, which can be realized using Eqs.~\eqref{scaledSSvarGen}, \eqref{eta} and \eqref{match},
\begin{equation} \label{zeta_Df}
    \zeta_{Df} = \frac{\lambda_{Df}}{\lambda_s}  = \frac{\sqrt{2}}{Q_f} = \frac{\Delta \zeta_{\rm sh}}{\sqrt{2}}
\end{equation}
Additionally, $\zeta_{Df}$ is also representative of the maximum scale of space charge separation in the ion fluid region, since the electron density in this region is greater than $N_{ef}$ $($i.e., $N_e(\zeta_c\leq\zeta<\zeta_f)>N_{ef})$. Thirdly, $Q_f$ and $|\zeta_c|$ govern the $\zeta-$scale of variation of the ion fluid profile $\Delta\zeta_i$. Since the expanding ion fluid extending up to $\zeta_f=Q_f$ contains a mass of ions $|\zeta_c|$ $\big($normalized to $n_{i0}\lambda_s(t)\big)$, the ion variation scale is given by
\begin{equation} \label{dZetai}
    \Delta\zeta_i=\min\{|\zeta_c|,Q_f\}
\end{equation}
Thus, $Q_f$ and $|\zeta_c|$ determine the scales of the electron sheath $\Delta \zeta_{\rm sh}$, electron-ion charge separation $\zeta_{Df} = \sqrt{2}\Delta\zeta_{\rm sh}$, and ion fluid variation $\Delta\zeta_i$.

To gain further insight into the aspects of the different dynamical regimes and delineate the transition among them, we study the parametric variation of $Q_f$ with respect to $\eta$ and $|\zeta_c|=-\zeta_c$. The variation of $Q_f(\eta,|\zeta_c|)$ is shown by the color density plot, Fig.~\ref{fig:paramVarPlots}$(b)$, with some contours at fixed $Q_f$ values plotted with solid lines in black. The contour $Q_f=1$ (plotted with a blue solid line) serves as boundary in the parameter space on which the ion expansion front $x_f$ lies at an acoustic correlation distance $\lambda_s$ from the initial vacuum interface. A qualitative classification of the expansion dynamics can be obtained based on the values of the variation scales of the electrons ($\Delta\zeta_{\rm sh}$) and ions ($\Delta\zeta_i$), and $|\zeta_c|$. In Fig.~\ref{fig:paramVarPlots}$(b)$ and the schematic Fig.~\ref{fig:paramVarPlots}$(c)$, we plot the boundary $|\zeta_c| = \sfrac{1}{Q_f} = \sfrac{\Delta\zeta_{\rm sh}}{2}$ for $Q_f<1$ in orange, along which, the plasma length $L$ is equal to the charge separation scale. Along the $|\zeta_c|=\Delta\zeta_i$ boundary in purple, the plasma length is equal to the ion variation length scale. These boundaries along with the $Q_f=1$ contour segment the parameter space domain into five regimes, as illustrated in Fig.~\ref{fig:paramVarPlots}$(c)$. The points in the parameter space for which the solutions are plotted in Fig.~\ref{fig:paramVarPlots}$(a)$ are denoted with green dots in Fig.~\ref{fig:paramVarPlots}$(c)$. These five regimes exhibit qualitatively distinct asymptotic plasma behaviors as described below.

In the low$-Q_f$ regimes~$(I) - (III)$, Eq.~\eqref{zeta_Df} indicates that the characteristic Debye length in the perturbed plasma region is greater than $\lambda_s(t)$. In the limit $\eta \ll 1$, the marginal response scale of the ions to the weak fields is given by $\Delta\zeta_i = Q_f \ll |\zeta_c|$. The asymptotic dynamics in regime~$(I)$ with $|\zeta_c| \gg \sfrac{1}{Q_f} \sim \zeta_{Df}$, obeys the ordering $L(t) \gg \lambda_{Df}(t) \gg \lambda_s(t)$. Thus, the nearly unperturbed ion slab shields the electrostatic field near the initial vacuum interface $x=0$ in a characteristic distance much smaller than the length of the self-similar plasma region. In the low$-Q_f$, low$-|\zeta_c|$ regimes~$(II)$ and $(III)$, on the other hand, the Debye length $\zeta_{Df}$ is much larger than the self-similar ion-fluid domain $\zeta_f-\zeta_c$, which leads to the formation of a diffuse electron plasma near the interface. For $|\zeta_c|\ll1$, the electron cloud has a density much lower than the unperturbed plasma density, and the ions remain largely unshielded. Assuming $N_{e\rm sh}\ll1$, Eqs.~\eqref{sheath-zeta} and \eqref{match} can be used to approximate the weakly shielded electrostatic field in regimes~$(II)$ and $(III)$
\begin{equation} \label{Qf-2,3}
    Q_f^{\left\{II,III\right\}} \simeq \eta |\zeta_c| - \frac{{Q_f^{\left\{II,III\right\}}}^2}{2}\left(Q_f^{\left\{II,III\right\}}+|\zeta_c|\right)
\end{equation}

Higher $Q_f$ values with increasing $\eta$ illustrate the stronger accelerating fields produced at higher $\eta$. $Q_f=P_f$ entails that in the $Q_f>1$ solutions the ions are accelerated to supersonic ion speeds at the leading ion edge, i.e., $v_{if}(t)>C_s(t)$. In contrast, the ion flows on the left of the $Q_f=1$ contour in blue remain subsonic throughout the expanding ion region with $P_f<1$. The strong acceleration experienced by the small mass of ions in the regimes~$(III)$ and $(IV)$ lying below the purple contour in Fig.~\ref{fig:paramVarPlots}$(b)$ lead to diffuse ion profiles varying rapidly over the scale
\begin{equation} \label{dZetai_3,4}
    \Delta \zeta_i^{\{III,IV\}} = |\zeta_c|
\end{equation}
As observed in Fig.~\ref{fig:paramVarPlots}$(a)$, the ion density in these regimes exhibit a steep density decrease near $\zeta_c$, followed by a low density ``tail" region extending up to $\zeta_f = Q_f>|\zeta_c|$ with subsonic and supersonic exit in regimes~$(III)$ and $(IV)$ respectively. In the high$-\eta$ regimes~$(IV)$ and $(V)$ with $Q_f\gtrsim1$, it can be shown using Eqs.~\eqref{zeta_Df} and \eqref{dZetai} that the maximum Debye length in the expanding ion fluid domain is smaller than the length of this domain. This gives rise to charge separation-induced electrostatic field oscillations at the Debye scale that modulate the ion density and velocity profiles in these regimes, as observed in the high$-\eta$ profiles in Fig.~\ref{fig:paramVarPlots}$(a)$. Lastly, in the high$-\eta$, high$-|\zeta_c|$ limit, the expansion in regime~$(V)$ is almost quasineutral throughout the self-similar domain with electron-ion charge separation $\mathcal{O}\left(\sfrac{1}{\eta}\right)$, due to Eq.~\eqref{SSuni:Poisson}.

The insensitivity of the plasma profiles with increasing $|\zeta_c|$ observed in Fig.~\ref{fig:paramVarPlots}$(a)$ leads to the $Q_f$ contours becoming vertical in Fig.~\ref{fig:paramVarPlots}$(b)$ at large values of $|\zeta_c|$. In regime~$(V)$, the correlation length $\lambda_s$ limits the spatial extent of the region in which the plasma profiles deviate from their unperturbed plasma state. In regime~$(I)$, on the other hand, the electron fluid shields the field over a characteristic $\lambda_{Df}$ scale. Thus, for large values of $|\zeta_c|$ the profiles near $\zeta_c$ approach the unperturbed plasma state in regimes~$(I)$ and $(V)$, consequently saturating the field \(Q_f\) with increasing \(|\zeta_c|\) for a fixed \(\eta\). $\lambda_s(t)$ and $\lambda_{Df}(t)$ are the two natural length scales for the variation of the electrostatic field driving the dynamics, corresponding to $\zeta-$scales of $1$ and $\zeta_{Df}$ respectively. Thus, the deviations of the profiles from the unperturbed plasma state are confined to $\max\{\zeta_{Df},1\}$ in the high$-|\zeta_c|$ limit, making the dynamics independent to $\zeta_c$.

In this section we have identified five dynamical regimes arising from the interplay of the relative magnitudes of $\lambda_D(t)$, $\lambda_s(t)$ and $L(t)$. In the next 2 sections, we discuss the asymptotic plasma dynamics in these $5$ regimes in detail. The quantities associated with regime $n\in\{I,II,III,IV, V\}$ are  denoted by a superscript $(n)$, and terms of the order $\mathcal{O}\left(\epsilon\ll1\right)$ are written in bold. The derived length scales for the dynamics are expressed in terms of the time-dependent length scales $L=L(t)$, $\lambda_s=\lambda_s(t)=\lambda_{s0}\exp{\left(\gamma t\right)}$ and $\lambda_D=\lambda_{D0}\exp{\left(\gamma t\right)}$.

\section{Fast heating (Low $\eta$)}\label{sec:lowEta}
When the electrons in a plasma of a modest density are heated rapidly such that $\gamma \omega_{pi0}^{-1}\gg 1$, the ions remain effectively stationary during the electron dynamics in the timescale $\gamma^{-1}$. The electrons produce an expanding boundary layer near the vacuum interface, whose properties are governed by this timescale and the mass of the heated electrons. The qualitative nature of the electron dynamics in this layer is characterized by the extent of charge separation. Since $L$ is proportional to $\lambda_D$, the mass of the heated electrons in the expansion region increases proportionally to the length scale of charge separation. With $\sfrac{L}{\lambda_D}$ large in regime~$(I)$, the electrons undergo Debye shielding dynamics in the boundary layer, while in regime~$(II)$ all of the heated electrons originating from $L\ll\lambda_D$ show a large global charge separation from the ions. The transition between the two regimes occurs when $L\approx\lambda_D$, or using the definitions~\eqref{eta} and \eqref{zeta_c-exp}, when $|\zeta_c| \approx \sfrac{1}{\sqrt{\eta}}$.

\subsection{Regime~I :  $\eta\ll1$, $|\zeta_c|\gg\sfrac{1}{\sqrt{\eta}}$}
\begin{figure}[!hbtp]
    \centering
    \includegraphics[width=\linewidth]{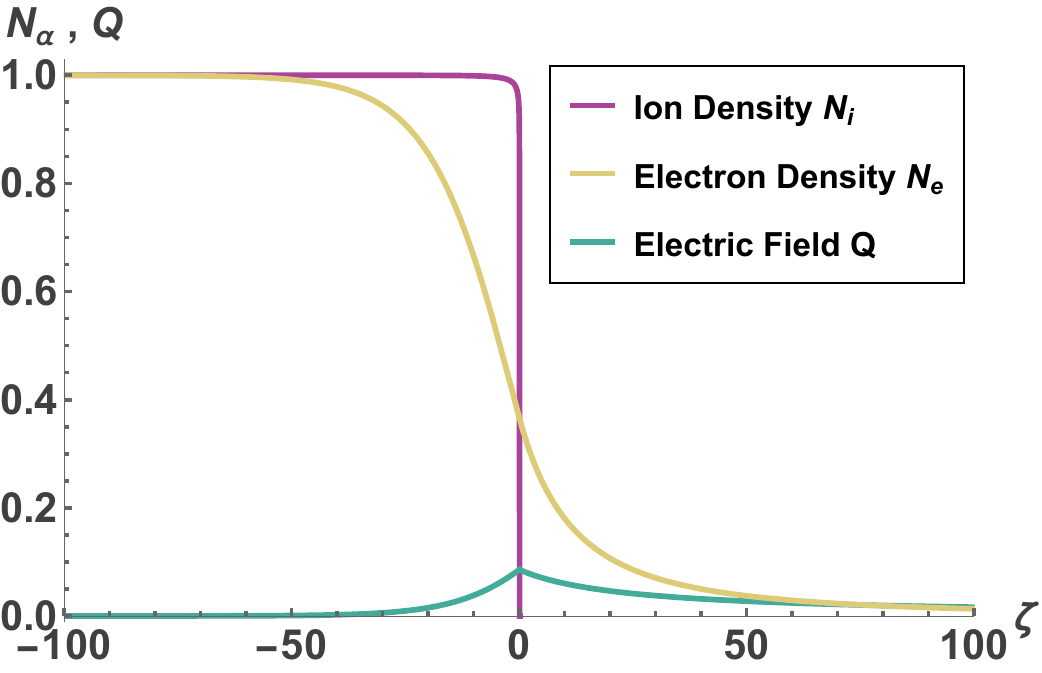}
    \caption{Ion and electron density profiles for  $\eta=10^{-2}$ and $\zeta_c=-100$ are plotted in purple and yellow respectively. The shielded electrostatic field is shown in green. The ions vary on the $\zeta-$scale $\Delta\zeta_i = Q_f\approx 8.578\times 10^{-2}$ while electron density and the field vary on the scale $\zeta_{Df} \approx 16.49$}
    \label{fig:leHzPlot}
\end{figure}

The plasma behavior in this regime is exemplified by the plasma density and electric field profiles in Fig.~\ref{fig:leHzPlot} for $\eta=10^{-2}$ and $\zeta_c=-100$. The Debye shielding dynamics of the electron fluid in an expanding boundary layer near $x=0$, and the nearly unperturbed ion fluid occupying $x<0$ is clearly evidenced. The length scales of ion fluid perturbation $\Delta L_i$, and of the shielding dynamics of the electrons $\Delta L_e$, are the essential lengths characterizing the dynamics. The saturation dynamics in regime~$(I)$ discussed in the previous section implies that the profiles in this case effectively represent the dynamics of a uniformly heated semi-infinite plasma slab. In practice, when $\lambda_D(t)$ is approximately proportional to $\lambda_s(t)$, and $L \gg \lambda_{D} \gg \lambda_{s}$, the plasma dynamics is independent of the length of the hot plasma region $L$, and can be approximated by the solutions in this regime.

We approximate the solutions in this regime by expanding the ion density $N_i^{(I)} = 1 + \boldsymbol{N_i}^{(I)}$, the velocity $P_i^{(I)} = \boldsymbol{P_i}^{(I)}$ and the field $Q^{(I)} = \boldsymbol{Q}^{(I)}$. Here, the perturbation of the ion density $\boldsymbol{N_i}^{(I)}=\mathcal{O}(\eta)$, and since the electron dynamics occurs over a large $\zeta - $scale of order $O(\sfrac{1}{\sqrt{\eta}})$, $\boldsymbol{P_i}^{(I)},\boldsymbol{Q}^{(I)} = \mathcal{O}(\sqrt{\eta})$. Then, Eqs.~\eqref{SSuni:sys} take the approximate form,
\begin{subequations} \label{Le_Hz_Eqns}
        \begin{align}
        & -\zeta\frac{d\boldsymbol{N_i}^{(I)} }{d\zeta} +  \frac{d \boldsymbol{P_i}^{(I)} }{d\zeta} = 0 \\
        & \boldsymbol{P_i}^{(I)}  -\zeta\frac{d \boldsymbol{P_i}^{(I)} }{d\zeta} -\boldsymbol{Q}^{(I)} = 0 \\
        & \frac{dN_{e}^{(I)}}{d\zeta} + N_{e}^{(I)} \boldsymbol{Q}^{(I)} = 0 \label{Le_Hz_Eqns:emom} \\
        & \frac{d\boldsymbol{Q}^{(I)}}{d\zeta} = \eta (1-N_{e}^{(I)}) \label{Le_Hz_Eqns:Poisson}
    \end{align} 
\end{subequations}
Using $\lim_{Q^{(I)}\to0} N_e^{(I)}\mkern-4mu \to \mkern-2mu 1$ along with Eqs.~\eqref{Le_Hz_Eqns:emom} and \eqref{Le_Hz_Eqns:Poisson}, $ \boldsymbol{Q}^{(I)}$ can approximated in terms of $N_e^{(I)}$ as
\begin{equation} \label{NeII-QII}
         \boldsymbol{Q}^{(I)} = \sqrt{2\eta\left(N_e^{(I)}-\ln N_e^{(I)}-1\right)}
\end{equation}
Imposing the matching condition~\eqref{match} provides the normalized electron density and sheath field at $\zeta_f$,
\begin{subequations}
            \begin{align}
                N_{ef}^{(I)} &\approx \exp(-1) \\
                Q_{f}^{(I)} &\approx \sqrt{2\eta} \exp{\left(-\frac{1}{2}\right)}
            \end{align}
\end{subequations}

The above solutions provide quantitative estimates for the sheath field and density, and for the length scales of electron and ion dynamics in this regime. The electron density in the sheath and in the ion fluid region varies over the $\zeta-$scale $\zeta_{Df} = \sfrac{\sqrt{2}}{Q_f} \approx \exp{\left(\frac{1}{2}\right)}\sfrac{1}{\sqrt{\eta}}$, or on the length scale given by
\begin{equation} \label{regI_DLe}
    \Delta L_e \approx \sqrt{\frac{\rm e}{\eta}}\lambda_s = \sqrt{\rm e}\lambda_{D}
\end{equation}
The ions near the origin experience a slight expansion on the length scale
\begin{equation} \label{regILi}
    \Delta L_i \approx \sqrt{\frac{2\eta}{\rm e}}\lambda_s =  \sqrt{\frac{2}{\rm e}}\frac{\lambda_{s}^2}{\lambda_{D}}~,
\end{equation}
obtained using Eq.~\eqref{dZetai}. These length scales being independent of $L$ explains the insensitivity of the profiles to $L$ (or $\zeta_c$) in this regime. Substituting $\eta = 10^{-2}$ in the above equations give the approximate normalized $\Delta L_i$ and $\Delta L_e$ values of $8.578\times 10^{-2}$ and $16.49$ respectively, for the profiles in Fig.~\ref{fig:leHzPlot}.

\subsection{Regime II: $\eta\ll1$, $|\zeta_c|\lesssim\sfrac{1}{\sqrt{\eta}}$}
\begin{figure} [!hbtp]
    \centering
    \includegraphics[width=\linewidth]{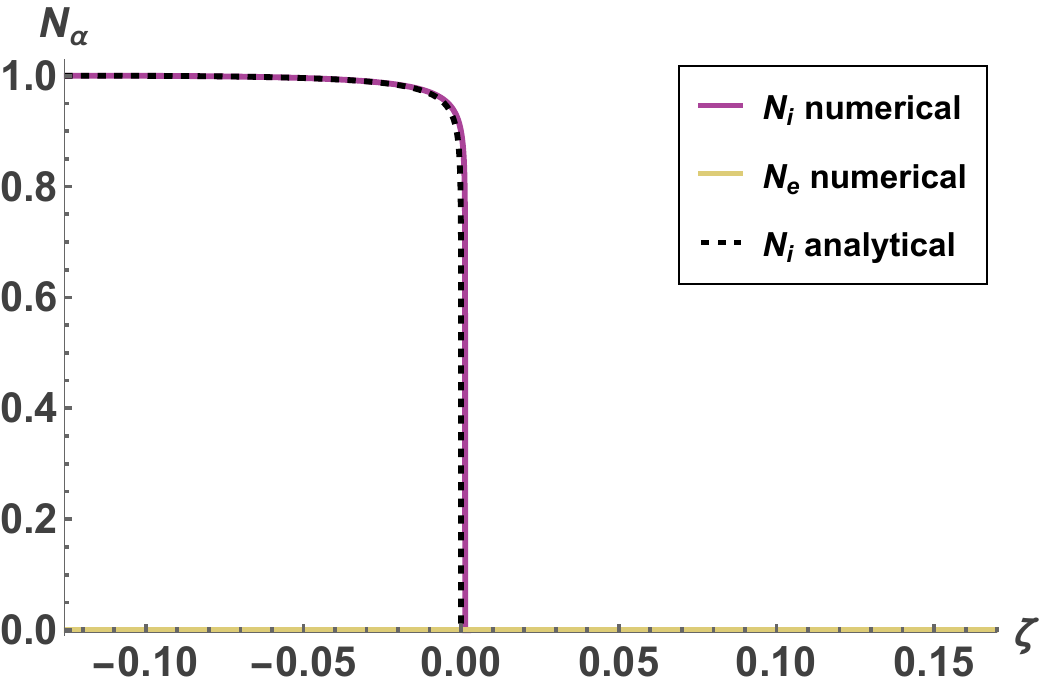}
    \caption{Ion density profile (Purple) and electron density profile (Yellow) for  $\eta = 10^{-2}$ and $\zeta_c = -0.126$. The black, dashed line is the analytic approximation, Eq.\eqref{Le_Lz_Approx_Ni}, for the ion density profile.}
    \label{fig:LeLzPlot}
\end{figure}

In Fig.~\ref{fig:paramVarPlots}$(a)$, the profiles in violet ($\eta=0.01$) in the first 4 columns represent the asymptotic dynamics in the limit of low$-\eta$ and low$-\eta \zeta_c^2$, as indicated by Fig.~\ref{fig:paramVarPlots}$(c)$. In this regime, a small mass of electrons in a boundary layer near the initial plasma-vacuum interface are rapidly heated on a timescale $\sim\gamma^{-1}$. These hot electrons originating from a region of length $L \lesssim \lambda_D$ are expelled from the boundary layer leaving a low electron density in this region. In the limit $L \ll \lambda_D$, an almost bare, unperturbed ion slab of dimension $L$ is left behind, that serves as a precursor for subsequent Coulomb explosion at later times. The electron and ion density profiles for $\eta = 10^{-2}$ and $\zeta_c = -0.126$ are shown in Fig.~ \ref{fig:LeLzPlot}. After a time $t$ of the heated electron dynamics, the plasma state is characterized by the ion density profile, and the characteristic values of the electron density and the unshielded electrostatic field.

The density profiles in this regime can by approximated by substituting the expansion $N_i^{(II)} = 1 + \boldsymbol{N_i}^{(II)}$, $P_i^{(II)} = \boldsymbol{P_i}^{(II)}$ and, $Q^{(II)} = \boldsymbol{Q}^{(II)}$ in Eqs.~\eqref{SSuni:sys}, where $\boldsymbol{N_i}^{(II)}, \boldsymbol{P_i}^{(II)}, \boldsymbol{Q}^{(II)} = \mathcal{O}(\eta)$. The resulting system can be solved to obtain the approximate dynamics in this limit for $\zeta<0$,
\begin{subequations} \label{Le_Lz_Sols}
        \begin{align}
        N_i^{(II)} &= 1 - \frac{\eta\left(1-N_{e0}\right)}{2} \left(\ln \left(\sfrac{\zeta}{\zeta_c}\right)\right)^2 \\
        P_i^{(II)} &= \eta\left(1-N_{e0}\right) \left(\zeta-\zeta_c-\zeta\ln \left(\sfrac{\zeta}{\zeta_c}\right)\right) \\
        N_e^{(II)} &= N_{e0} \left(1-\eta\left(1-N_{e0}\right)\frac{\left(\zeta-\zeta_c\right)^2}{2}\right) \\
        Q^{(II)} &= \eta\left(1-N_{e0}\right)\left(\zeta-\zeta_c\right)
    \end{align} 
\end{subequations}
Since the electron density varies over a distance $\lambda_{Df}$ that is larger than the length $L$ of the expanding boundary layer, the electron density is nearly uniform in this layer, $N_e(\zeta_c < \zeta < \zeta_f) \approx N_{e0}\approx N_{ef}$. Since the ion fluid variation scale is much smaller than the boundary layer length ($\zeta_f \ll |\zeta_c|$), we obtain $Q_f\approx -\eta\left(1-N_{e0}\right)\zeta_c$ by setting $\zeta_f \approx 0$ in Eqs.~\eqref{Le_Lz_Sols}. Then, condition \eqref{match} gives the characteristic electron density in the boundary layer,
\begin{equation} \label{lowEtaNe0}
            N_{e0} \simeq 1 + \frac{1-\sqrt{1+2\eta\zeta_c^2}}{\eta \zeta_c^2}
\end{equation}
This can be further approximated in the $\eta\zeta_c^2\ll1$ limit to give
\begin{subequations}
\begin{align}
    N_{e0}^{(II)} &\approx N_{ef}^{(II)} \approx \frac{\eta \zeta_c^2}{2} \label{lowEtaNe0Approx} \\
    Q_{f}^{(II)} &\approx |\eta \zeta_c| \label{Qfapprox-2}
\end{align}
\end{subequations}
and the approximate solutions,
\begin{subequations} \label{Le_Lz_SolsApprox}
        \begin{align}
        N_i^{(II)} &\simeq 1 - \frac{\eta}{2} \left(\ln \left(\sfrac{\zeta}{\zeta_c}\right)\right)^2 \label{Le_Lz_Approx_Ni} \\
        P_i^{(II)} &\simeq \eta \left(\zeta-\zeta_c-\zeta\ln \left(\sfrac{\zeta}{\zeta_c}\right)\right) \\
        N_e^{(II)} &\simeq \frac{\eta \zeta_c^2}{2} \left(1-\frac{\eta}{2}\left(\zeta-\zeta_c\right)^2\right) \\
        Q^{(II)} &\simeq \eta\left(\zeta-\zeta_c\right)
    \end{align} 
\end{subequations}
where condition~\eqref{match} has been used. The analytical approximation for the ion density, Eq.~\eqref{Le_Lz_Approx_Ni} is shown with the dashed line in Fig.~\ref{fig:LeLzPlot}.

Equation~\eqref{Qfapprox-2} is the correct asymptotic limit of Eq.~\eqref{Qf-2,3} for $Q_f \!\ll \!1$, and explains the approximately straight line contours for $Q_f$ in this limit in the $\log{\eta}-\log{|\zeta_c|}$ plane in Fig.~\ref{fig:paramVarPlots}$(b)$. The electron and ion density variation scales are provided by Eqs.~\eqref{zeta_Df}, \eqref{dZetai} and \eqref{Qfapprox-2} as $\Delta\zeta_{\rm sh} = \sfrac{2}{\eta |\zeta_c|}$ and $\Delta\zeta_i = \eta |\zeta_c|$. In other words, the diffuse electron density varies over a length scale
\begin{equation}
    \Delta L_s = \frac{2}{\eta |\zeta_c|} \lambda_s = \frac{2\lambda_D^2}{L} ~,
\end{equation}
while the ion slab has a minimal perturbation near the origin over a distance of the order
\begin{equation}
    \Delta L_i = \eta |\zeta_c| \lambda_s = \frac{\lambda_s^2}{\lambda_D^2} L
\end{equation}
If the heating ceases at a time $t$, the ion slab of dimension $L(t)$ with a very low density of electrons $n_{e0}\sfrac{\eta\zeta_c^2}{2}$, becomes susceptible to Coulomb explosion.

\section{Slow heating (High $\eta$)} \label{sec:highEta}
At higher plasma densities or slower rates of electron heating with $\omega_{pi0} \gg \gamma$, the ion response to the electrostatic field (varying on a timescale $\gamma^{-1}$) is rapid. While the electron heating slowly changes the thermal pressure, the ions respond to restore the pressure balance via acoustic/rarefaction motions. In the high$-\eta$ regimes $(IV)$ and $(V)$, an ion wave expands to a distance on the order of the ion correlation length $\lambda_s$. Thus, we observe $\zeta_f = Q_f =\mathcal{O}(1)$ for the high$-\eta$ ion profiles in Fig.~\ref{fig:paramVarPlots}$(a)$, and in the $Q_f$ contours in these regimes in Fig.~\ref{fig:paramVarPlots}$(b)$. Since these regimes lie to the right of the $Q_f=1$ contour in Fig.~\ref{fig:paramVarPlots}$(b)$, the ions at the leading edge of the expansion wave exit at a supersonic speed. For $|\zeta_c|\gg 1$, the  $Q_f = 1$ contour vertically asymptotes at $\eta \approx 3.115$. On the other hand, Eq.~\eqref{Qf-2,3} leads to the expression $\eta |\zeta_c| \approx \sfrac{3}{2}$ for the asymptotic behavior of this contour as $\zeta_c \rightarrow 0$. Thus, with increasing $\zeta_c$ the dynamics transitions from regime~$(III)$ when $|\zeta_c| \ll \sfrac{3}{2\eta}$, to regime $(IV)$ for $\sfrac{3}{2\eta} \ll|\zeta_c| \ll 1$, and to regime~$(V)$ for $|\zeta_c| \gtrsim 1$.

Since the electron inertia is neglected, the influence of electrons on the qualitative nature of the expansion is governed by the extent of charge separation in the perturbed plasma region. Due to Eq.~\eqref{eta}, $\eta$ determines the variation scale of electron-ion charge separation relative to $\lambda_s$. Since $\zeta_f \gg \zeta_{Df}$ in the high$-\eta$ regimes~$(IV)$ and $(V)$, the plasma profiles near $\zeta_f$ in regime~$(IV)$ and over the whole ion fluid domain in regime~$(V)$ are close to quasineutral. The electric field profiles in these regions in Fig.~\ref{fig:paramVarPlots}$(a)$ exhibit charge separation induced Debye scale oscillations. These oscillatory features represent forward and backward propagating electrostatic waves for $\zeta>0$ and $\zeta<0$ respectively. As the electron density $N_e$ approaches $N_{ef}$ near $\zeta_f$, the oscillations of the ion density profiles around $N_{ef}$ produce collisionless shock like structures near $\zeta_f$, as noticed by an increase in the ion density near $\zeta_f$ in Fig.~\ref{fig:paramVarPlots}(a). The mean plasma behavior on which the oscillatory features are superposed are given by the quasineutral solutions in these regions. The approximate mean plasma profiles to leading order in $\sfrac{1}{\eta}$  in these regimes can be obtained by setting $N_i\approx N_e=\bar{N}$ in Eqs.~\eqref{SSuni:sys}, and using Eqs.~\eqref{SSuni:cont} - \eqref{SSuni:emom} to find the differential equation for the mean plasma velocity $\bar{P}$,
\begin{equation} \label{SSmean:P}
         \bar{P} + \left(\bar{P}-\zeta-\frac{1}{\bar{P}-\zeta}\right)\frac{d\bar{P}}{d\zeta} = 0 
\end{equation}
The mean field $\bar{Q}$ and density $\bar{N}$ obey
\begin{equation} \label{SSmean:Q-N}
        \bar{Q} = -\frac{1}{\bar{N}}\frac{d\bar{N}}{d\zeta} = \frac{1}{\bar{P}-\zeta}\frac{d\bar{P}}{d\zeta}
\end{equation}
where the overbars denote the mean quantities.

The mass of the perturbed plasma $|\zeta_c|$ is the second factor influencing the magnitude of the field $Q_f$, and the qualitative response of the ions. A low mass of electrons in the $|\zeta_c| \ll 1$ limit produces a low density, diffuse electron cloud within the expanding ion wave. In regimes~$(III)$ and $(IV)$, the effect of the small electron mass on the electron fluid behavior is identical to that in regime~$(II)$. The small ion mass in these regimes originates from a length $L(t)\ll\lambda_s(t)$ of the initial plasma, and thus produces a rarefied ion wave characterized by ion density tails as observed in Fig.~\ref{fig:paramVarPlots}$(a)$. In regimes~$(III)$ and $(IV)$, the rapid variation of the ion density from its unperturbed value at $\zeta_c$ occurs over the scale $|\zeta_c|$, or on a length scale
\begin{equation} \label{dli_3,4}
    \Delta L_i^{\{III,IV\}} = L
\end{equation}
Analytical approximations for the rapidly varying ion profiles near the initial plasma-vacuum interface are derived in appendix~\ref{App:Prof34}. For the ion density and velocity we obtain the approximate relations
\begin{subequations} \label{ion_3,4_near0}
    \begin{align}
        N_i &= \left|\zeta_c\right| \left(|\zeta|\left(B^{\frac{1}{3}} + \sgn(\zeta) B^{-\frac{1}{3}}\right)^2 - \zeta\right)^{-1} \label{den_3,4_near0} \\
        P_i &= |\zeta|\left(B^{\frac{1}{3}} + \sgn(\zeta) B^{-\frac{1}{3}}\right)^2
    \end{align}
\end{subequations}
in this region. Here $\sgn$ is the sign function, and the expression for $B(\eta,\zeta_c)$ is given by Eq.~\eqref{BVal}.

\subsection{Regime III :  $\eta\gg1$, $|\zeta_c|\ll  \sfrac{3}{2\eta}$}
\begin{figure} [!hbtp]
    \centering
    \includegraphics[width=1\linewidth]{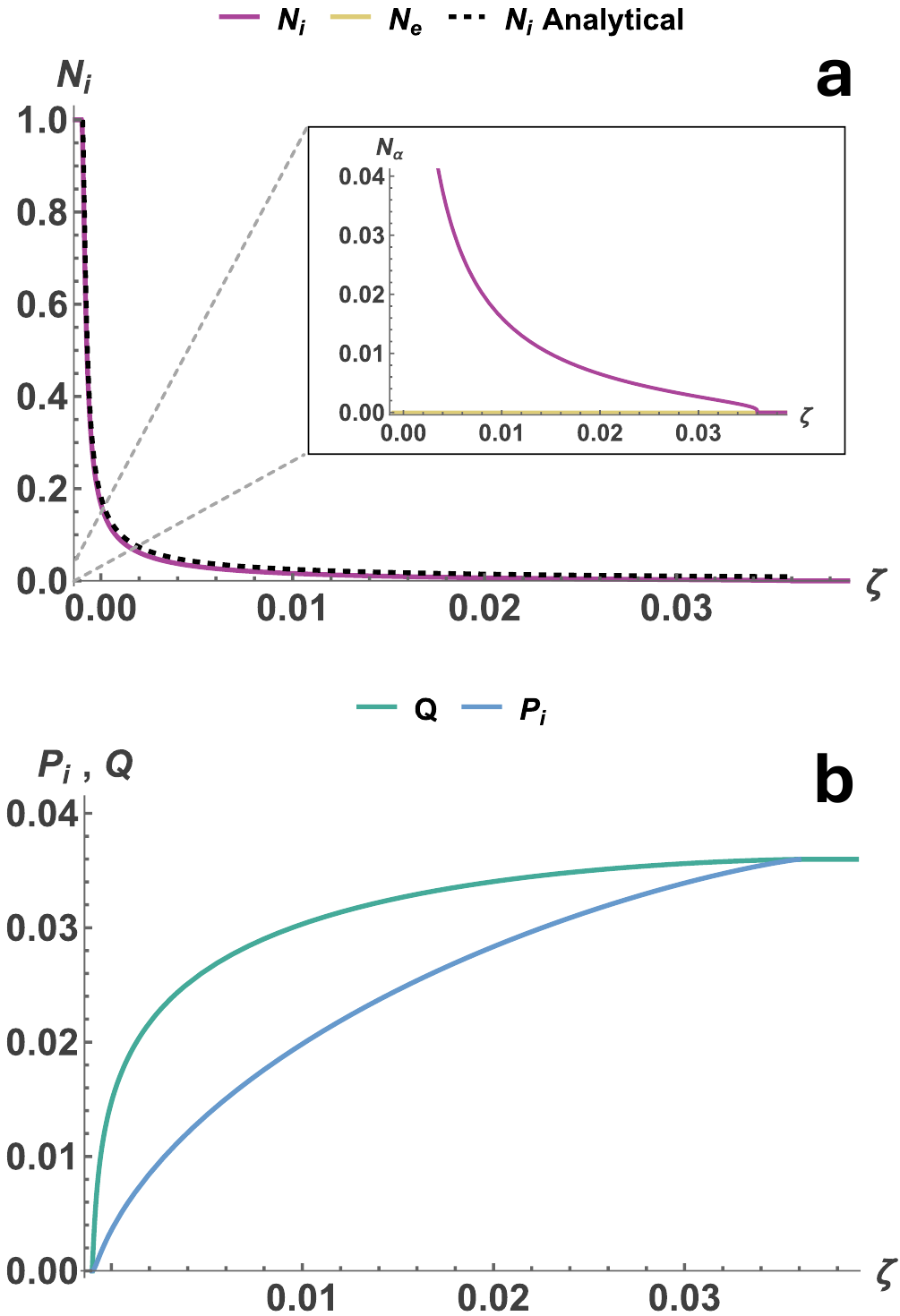}
    \caption{Plasma profiles for $\eta=36$ and $\zeta_c=-10^{-3}$ : $(a)$ Ion density profile in purple solid line, and its analytical approximation near $\zeta_c$ and origin (Eq.~\eqref{den_3,4_near0}) in black dashed line; $N_i$ and $N_e$ in the ion density tail region are shown in the inset plot. $(b)$ Ion velocity (blue solid line) and electric field (green solid line) profiles}
    \label{fig:Reg3Plots}
\end{figure}

A representative dynamics with $\eta=36$ and $\zeta_c=-10^{-3}$ is plotted in Fig.~\ref{fig:Reg3Plots}, that illustrates the steep ion density decrease near $\zeta_c$ and the low density tail. Figure~\ref{fig:Reg3Plots}$(a)$ shows the complete ion density profile, along with the analytical approximation for $N_i$ in Eq.~\eqref{den_3,4_near0}. The electron and ion densities in the tail region are plotted on an enlarged scale in the inset plot, where the nearly uniform, diffuse electron profile can be seen at a much lower density than the ions. Similar to regime~$(II)$, the electrons in this regime are also almost completely evacuated from heated plasma region leaving an ion fluid slab. Thus, in effect, the ion dynamics in this regime represents the onset of Coulomb explosion of a thin ion slab of dimension $L$. For the impending Coulomb explosion following the termination of electron heating, the key quantities of interest from the self-similar solutions are the electron and ion density profiles and the dimension of the slab.

Equation~\eqref{Qf-2,3} with $Q_f^{(III)}\ll1$ gives the value of the nearly uniform electron density $N_{e0}^{(III)} \approx N_{ef}^{(III)} \approx \sfrac{\eta \zeta_c^2}{2}$. Since Eq.~\eqref{Qfapprox-2} is also applicable here, we obtain the expression for the position and velocity of the ion expansion front, $\zeta_f = P_f \approx \eta|\zeta_c|$. The analytical expression for the ion density derived in Appendix~\ref{App:Prof34} provides a good approximation for the ion fluid behavior near $\zeta_c$ and the origin. Due to assumption~\eqref{near0_ass}, this approximation is valid for $\zeta \gtrsim \zeta_c \left(1 - \sfrac{2}{\eta}\right)$. Since $\eta \gg 1$, these solutions give an accurate description for the density drop from the unperturbed plasma as noticed in Fig.~\ref{fig:Reg3Plots}$(a)$. For a simple estimate of this density drop near $\zeta_c$, a cruder approximation for the profiles can be obtained by assuming $P_i\ll \zeta,Q$, leading to the relations
\begin{subequations} \label{EqnsNearZetac}
    \begin{align}
       Q &=-\zeta \dfrac{dP_i}{d\zeta }\\
       N_i\dfrac{dP_i}{d\zeta} &=\zeta \dfrac{dN_i}{d\zeta } \\
       \dfrac{dQ}{d\zeta} &= \eta N_{i}
    \end{align}
\end{subequations}
These can then be solved using Eqs.~\eqref{coreBCs} to give
\begin{subequations} \label{nearZetac_3,4}
      \begin{align}
     N_{i} &\approx \left(\frac{\zeta}{\zeta_{c}} \right)^{\sqrt{\eta}-2} \label{NiReg3} \\
     P_i &\approx \left( \sqrt{\eta} - 2\right) \left( \zeta -\zeta _{c}\right) \\
     Q &\approx \zeta \left[2\eta\left(1-\left(\frac{\zeta}{\zeta_{c}} \right)^{\sqrt{\eta}-2}\right)\right]^{\frac{1}{2}}
      \end{align}
\end{subequations}
for $-\zeta > -\zeta_c \left(1-\sfrac{1}{\sqrt{\eta}}\right)$. Equation~\eqref{NiReg3} suggests that the steep density decrease takes place on a scale $\sfrac{|\zeta_c|}{\sqrt{\eta}}$, or on a length scale
\begin{equation} \label{RegIIIDLidrop}
    \Delta L_{i,\rm drop} = \frac{ |\zeta_c|}{\sqrt{\eta}} \lambda_s = \frac{\lambda_D L}{\lambda_s} \ll L
\end{equation}
Therefore, as the electrons are heated, an expanding, bare ion fluid slab of length
\begin{equation}
     L_i = \left(1 + \eta\right) |\zeta_c| \lambda_s = \left(1 + \frac{\lambda_s^2}{\lambda_D^2}\right) L
\end{equation}
is produced, with a highly non-uniform ion distribution that can be approximated by Eq.~\eqref{den_3,4_near0}. The electrons within the slab drop to a low density of $n_{e0}\frac{L^2}{2\lambda_D^2}$.

\subsection{Regime IV : $\eta\gg1$, $\sfrac{3}{2\eta} \ll|\zeta_c|\ll1$}
\begin{figure} [!hbtp]
    \centering
    \includegraphics[width=1\linewidth]{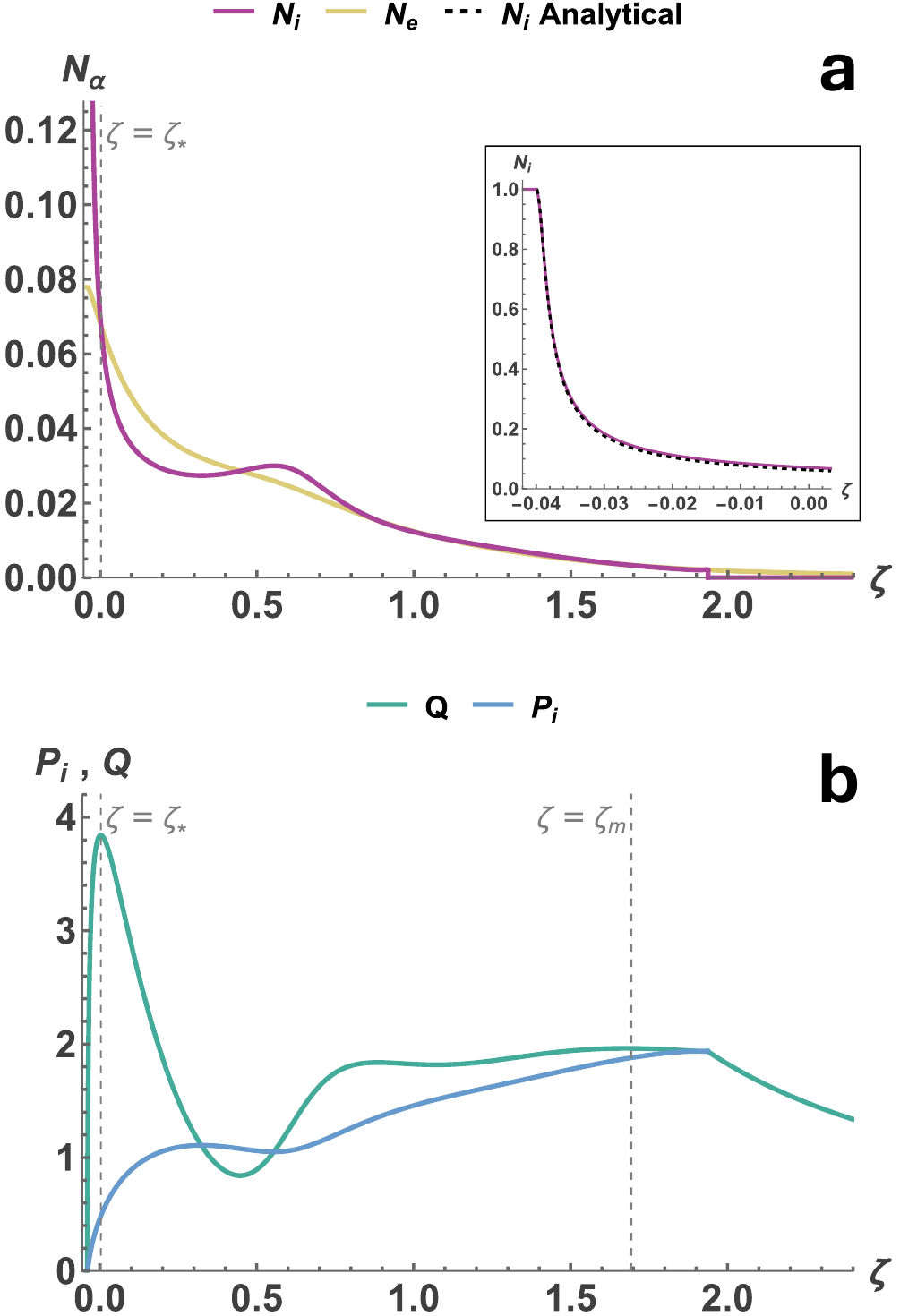}
    \caption{Plasma profiles for $\eta = 900$ and $\zeta_c=-0.04$ : $(a)$ Ion and electron density profiles; The ion dynamics (purple line) along with the analytical approximation (black dashed line) Eq.~\eqref{den_3,4_near0} are plotted until $\zeta_*$ in the inset plot. $(b)$ Ion velocity and electric field profiles; $\zeta=\zeta_m$ is shown in $(b)$, and $\zeta=\zeta_*$ is shown in both $(a)$ and $(b)$ with gray dashed lines.}
    \label{fig:Reg4Plots}
\end{figure}

In regime~$(IV)$, the ion density profiles exhibit an identical rapid decrease from the unperturbed density $n_{i0}$ near $\zeta_c$, as that in regime~$(III)$. The small mass of electrons spontaneously escaping from the hot, expanding plasma region into vacuum leads to a low electron density in this regime as well. The essential qualitative difference of the dynamics compared to regime~$(III)$ occurs on account of a small Debye length $\zeta_{Df} \ll \zeta_f$, that leads to a nearly quasineutral ion density tail. The rapid response of the ions to the strong mean fields produced in this rarefied expansion wave accelerates them to supersonic speeds. The plasma behavior in this regime can be noticed in the plots in Fig.~\ref{fig:Reg4Plots} for $\eta = 900$ and $\zeta_c=-0.04$. Due to the strong ion acceleration and a small mass of heated electrons near the initial plasma-vacuum interface, this regime exhibits a very high energy transfer from the externally heated electrons to the kinetic energy of the ions. The energy partitioning will be detailed in Sec.~\ref{sec:energetics}. 

To obtain analytical estimates for the profiles in this regime, we firstly note that the mass of the electrons in the sheath $\mathcal{N}_{es}$ is given by $\mathcal{N}_{es}=\sfrac{Q_f}{\eta}$ using Eq.~\eqref{sheath-Nes}. Thus, with $Q_f=\mathcal{O}(1)$, a large majority of the electrons originating from an initial plasma region of length $|\zeta_c|$ occupy a region of length $\approx \zeta_f=\mathcal{O}(1)$, leading to $N_{e0}^{(IV)}=\mathcal{O}(\left|\zeta_c\right|)$. Then, the transition from the ion density drop region near $\zeta_c$ to the tail region can be demarcated by the point $\zeta_*$, where the ion density drops to $\mathcal{O}\left(|\zeta_c|\right)$ and intersects the low electron density profile. As seen in Fig.~\ref{fig:Reg4Plots}$(a)$, the flow beyond beyond $\zeta_*$ turns increasingly quasineutral until $\zeta_f$. The behavior in the density drop region is captured well by Eqs.~\eqref{ion_3,4_near0} and \eqref{nearZetac_3,4}. In the inset plot in Fig.~\ref{fig:Reg4Plots}$(a)$, the density drop near $\zeta_c$ is shown until $\zeta_*$, along with the analytical approximation in this region given by Eq.~\eqref{den_3,4_near0}. Since $\sfrac{|\zeta_c|}{\sqrt{\eta}}$ is the scale for the ion density drop supplied by the approximation~\eqref{NiReg3}, the ion density drops faster with $\zeta$ for higher $\eta$, and $\zeta_*-\zeta_c$ is lower. The field reaches a maximum at $\zeta_*$, and exhibits electrostatic oscillations for $\zeta>\zeta_*$, leading to pronounced modulations in the ion velocity profile in the quasineutral tail. With a small charge separation $N_i - N_e = \mathcal{O}(\sfrac{1}{\eta})$, Eqs.~\eqref{SSmean:P} and \eqref{SSmean:Q-N} can be used to approximate the mean plasma dynamics in this nearly quasineutral tail region.

The high energy gained by the ions near the ion expansion front is a key feature of the dynamics in this regime. Near $\zeta_f$, the mean field $\bar{Q}$ reaches a local maximum at $\zeta_m$ where,
\begin{equation}\label{PatZetam}
   \bar{P}(\zeta_m) = \frac{1}{3}\left(2\zeta_m+\sqrt{3+\zeta_m^2}\right)
\end{equation}
At $\zeta_f$, $\bar{P}|_{\zeta_f}=\bar{Q}|_{\zeta_f}=\zeta_f$ due to the condition~\eqref{ZetafVals}, and the $\eta$ dependence of $\zeta_f$ is supplied by condition~\eqref{match}. Then, we obtain the following approximate profiles for the mean density and velocity near $\zeta_f$,
\begin{subequations} \label{solsnearzetaf}
\begin{align}
       \bar{N} &= \frac{Q_f^2}{2\eta}\exp{\left[Q_f^2(1-\sfrac{\zeta}{Q_f})\right]} \\
        \bar{P} &= Q_f\left(1-\frac{1}{2}\left(Q_f-\zeta\right)^2\right)
\end{align}
\end{subequations}
For $\zeta_m$ close to $\zeta_f$, Eqs.~\eqref{PatZetam} and \eqref{solsnearzetaf} can be used to approximate $\zeta_m$ in terms of $Q_f$ to leading order in $\left(Q_f\!-\!\zeta\right)^2$,
\begin{equation}\label{ZetamApprox}
   \zeta_m \approx 2Q_f-\sqrt{1+Q_f^2}
\end{equation}
This approximation for $\zeta_m$ has an error $\mathcal{O}\left(\left(Q_f-\zeta_m\right)^2\right)$, which decreases with increasing $Q_f$. $\left(Q_f-\zeta_m\right)^2= (\sqrt{2}-1)^2 \approx 0.17$ for sonic ion exit flows sets an upper bound to the value of $\left(Q_f-\zeta_m\right)^2$ for the higher $\eta$ values of regime~$(IV)$. Equation~\eqref{ZetamApprox} thus provides a good estimate for $\zeta_m$ for supersonic ion flows with an error of less than $20\%$. $\zeta = \zeta_*$ and $\zeta = \zeta_m$ are represented with gray dashed lines in the Fig.~\ref{fig:Reg4Plots}. The decrease in $\bar{Q}$ from $\zeta_m$ to $\zeta_f$ can be observed to cause a flattening of the ion velocity profile beyond $\zeta_m$ and $\frac{d\bar{P}}{d\zeta}\rightarrow0$ as $\zeta\rightarrow\zeta_f$. 

The ion density in this regime varies on a scale $\Delta L_i = L$, and drops rapidly from $N_i^{(IV)}=1$ at $\zeta_c$ to $\mathcal{O}(\zeta_c)$ near the initial vacuum plasma interface over a length scale
\begin{equation} \label{regIV_DLidrop}
    \Delta L_{i,\rm drop} = \frac{\lambda_{D}L}{\lambda_{s}}
\end{equation}
This is followed by a quasineutral tail region where the density decreases from $\mathcal{O}(|\zeta_c|)$ to $\mathcal{O}(\sfrac{1}{\eta})$ over a distance $\mathcal{O}(\lambda_s)$ from the origin.

\subsection{Regime V :  $\eta\gg1$, $|\zeta_c|\gtrsim1$}
\begin{figure*} [!hbtp]
    \centering
    \includegraphics[width=1\linewidth]{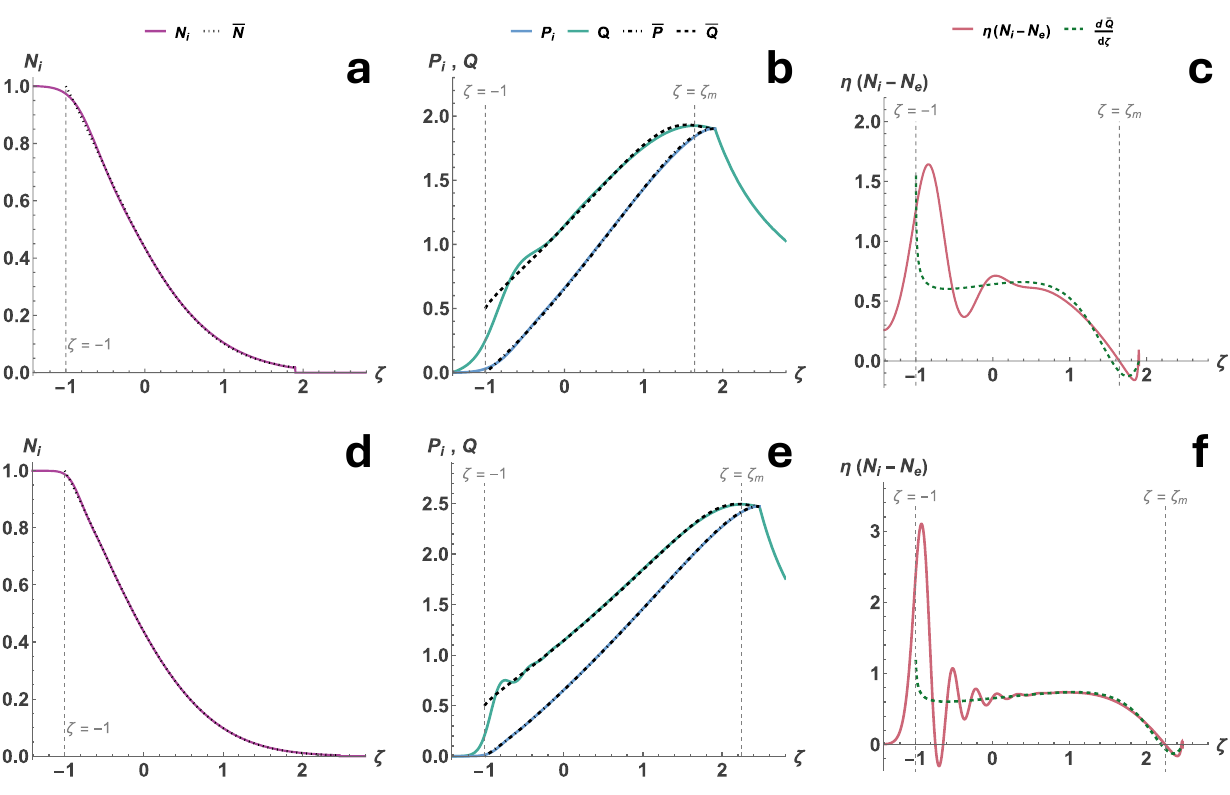}
    \caption{Ion density profiles for $(a)$ $\eta=100$ and $(d)$ $\eta=900$ with $\bar{N}$ plotted with black dotted lines. Ion velocity and electrostatic field profiles for $(b)$ $\eta=100$ and $(e)$ $\eta=900$ with $\bar{P}$ and $\bar{Q}$ plotted with black dot-dashed and dashed lines respectively. Plots $(c)$ and $(f)$ show the variation of $\eta(N_i-N_e)$ in red along with $\frac{d\bar{Q}}{d\zeta}$ in green dashed lines for $\eta$ values of $100$ and $900$ respectively.  $\zeta=-1$ and $\zeta=\zeta_m$ are represented with gray dashed lines the plots}
    \label{fig:HeHzplots}
\end{figure*}

The quasineutral expansion wave in regime~$(V)$ can be observed in Fig.~\ref{fig:HeHzplots}, where the profiles in this regime for $\eta$ values of $100$ and $900$ are plotted for comparison. The mean solutions obtained by solving Eqs.~\eqref{SSmean:P} and \eqref{SSmean:Q-N} are also shown. The solutions can be clearly noticed to asymptote to the unperturbed plasma state for $-\zeta_c\gtrsim 1$. This behavior is explained by the mean field extending up to a correlation distance $\lambda_s$ in the negative$-x$ direction from the initial plasma-vacuum interface. All desired solutions of Eq.~\eqref{SSmean:P} with velocity increasing from $0$ emerge from $\zeta=-1$. Thus, the length scale of expansion of the close to quasineutral plasma is given by
\begin{equation} \label{regV_DL}
    \Delta L_{e,i}(t) = \lambda_s(t) = \frac{C_{s0}}{\gamma} \exp{\left(\gamma t\right)}
\end{equation}

To further illustrate the insensitivity of the dynamics for $-\zeta_c\gtrsim 1$ for a fixed $\eta$, we approximate the mean dynamics near $\zeta=-1$ from Eqs.~\eqref{SSmean:P} and \eqref{SSmean:Q-N},
\begin{subequations} \label{barProfilesNearZetac}
        \begin{align}
            \bar{N} &= \frac{1}{2}\left(1-\zeta\right) \\
            \bar{P} &= \frac{1}{2}\left(1+\zeta\right) \\
            \bar{Q} &= \frac{1}{1-\zeta}
        \end{align}
\end{subequations}
The weak discontinuity of $\bar{N}$ and $\bar{P}$ at $\zeta=-1$ is a well known feature that appears at the boundary of the unperturbed plasma and the rarefaction wave in quasineutral self-similar solutions~\cite{SACK1987311,Schmalz_SS_type2}. This is accompanied by a weak discontinuity in the electrostatic potential $\Phi$ at this boundary, which shows up as a discontinuity of the mean electric field $\bar{Q}$ at $\zeta=-1$ in our mean solutions. This discontinuity is smoothed out in the complete solution of our system, when the small parameter $\sfrac{1}{\sqrt{\eta}}$ is retained through the Poisson equation. This fact has been pointed out in the literature as a characteristic qualitative difference of the solutions of the dispersive hydrodynamic system as compared to the Euler equations~\cite{DispersiveHydroShocksgurevich1984}. The transition from the unperturbed plasma profiles at $\zeta\lesssim-1$ to the rarefaction region takes place in a boundary layer, in which the profiles are governed by the equations
\begin{subequations} \label{He_Hz_boundary}
        \begin{align}
        \frac{d\tilde{N}_e}{d\tilde{\zeta}} = \left(-1+\tfrac{1}{\sqrt{\eta}}\tilde{\zeta}\right)\frac{d\tilde{P}_i}{d\tilde{\zeta}} &= \left(-1+\tfrac{1}{\sqrt{\eta}}\tilde{\zeta}\right)^2\frac{d\tilde{N}_i}{d\tilde{\zeta}} = -Q  \\
        \frac{dQ}{d\tilde{\zeta}} &= \tilde{N}_i-\tilde{N}_e
    \end{align} 
\end{subequations}
with $\tilde{\zeta}=\sqrt{\eta}(\zeta+1)$, $\tilde{N}_\alpha=\sqrt{\eta}\boldsymbol{N_\alpha}$ and $\tilde{P}_i=\sqrt{\eta}\boldsymbol{P_i}$. The solution for the field variation from Eqs.~\eqref{He_Hz_boundary} takes the form of an Airy function,
\begin{equation}
    Q \sim Ai\left[-\left(\frac{2}{\sqrt{\eta}}\right)^{\sfrac{1}{3}}\tilde{\zeta}\right] 
\end{equation}
which governs the transition. The boundary layer near $\zeta=-1$, thus, has a characteristic scale $\Delta\zeta_b=\eta^{\sfrac{-1}{3}}$. 

In the bulk of the rarefaction region, oscillations in plasma profiles about the mean solutions represent electrostatic waves generated due to charge separation. The oscillatory behavior in the bulk of the rarefaction region can be modeled by introducing a small scale variable $\tilde{\zeta}=\zeta/\epsilon$ with $\epsilon \sim \mathcal{O}\left(\sfrac{1}{\sqrt{\eta}}\right)$ and carrying out a two scale expansion of the system around the mean solutions. This procedure is detailed in appendix~\ref{regV2scale}, and the perturbations to the mean profiles on the order $\mathcal{O}(\sfrac{1}{\eta})$ are obtained, along with the superposed oscillations with a spatially varying wavenumber $k(\zeta)=\left[\eta\bar{N}\left(\tfrac{1}{(P-\zeta)^2}-1\right)\right]^{\sfrac{1}{2}}$. The charge separation $N_i-N_e$ along with the mean charge separation profiles obtained from the perturbative solutions can be seen in the example figures~\ref{fig:HeHzplots}$(c)$ and \ref{fig:HeHzplots}$(f)$. The oscillations get damped from $\zeta=-1$ to a region close to $\zeta=1$, corresponding to distances of $\lambda_s$ on either sides of $x=0$ in real space. Near $\zeta=0$, Eqs.~\eqref{SSmean:P} and \eqref{SSmean:Q-N} gives rise to the following approximate behaviors for $\bar{P}$ and $\bar{N}$
\begin{subequations}
    \begin{align}
        \bar{P} &= \bar{P}_0\left(1+\frac{P_0}{1-P_0^2}\zeta\right) \\
        \bar{N} &= \bar{N}_0\exp{\left[-\frac{P_0}{1-P_0^2}\left(\zeta+\frac{P_0^3}{2(1-P_0^2)}\zeta^2\right)\right]}
    \end{align}
\end{subequations}
with $\{\bar{P},\bar{N}\}|_{\zeta=0}=\{\bar{P}_0,\bar{N}_0\}$. Using these, we can obtain the approximate spatial variation of the oscillation wavenumber $k$ in the expansion bulk as
\begin{multline}
     k^2 = \eta \bar{N}_0 \left[\left(\bar{P}_0+\frac{2\bar{P}_0^2-1}{1-\bar{P}_0^2}\zeta\right)^{-2}-1\right] \\
     \times \exp{\left[-\frac{\bar{P}_0}{1-\bar{P}_0^2}\left(\zeta+\frac{\bar{P}_0^3}{2(1-\bar{P}_0^2)}\zeta^2\right)\right]}
\end{multline}
Lastly, beyond $\zeta=1$, the dynamics in regimes~$(IV)$ and $(V)$ become identical, and the $P_i$ profiles flatten beyond $\zeta_m$ where $\bar{Q}(\zeta_m)$ reaches a maximum. Since the analysis in the region between $\zeta_m$ and $\zeta_f$ carried out in the previous subsection is also valid here, Eq.~\eqref{ZetamApprox} also serves as an approximation for $\zeta_m$ in terms of $Q_f$ in this regime.

\begin{figure*}[!hbtp]
    \centering
    \includegraphics[width=1\linewidth]{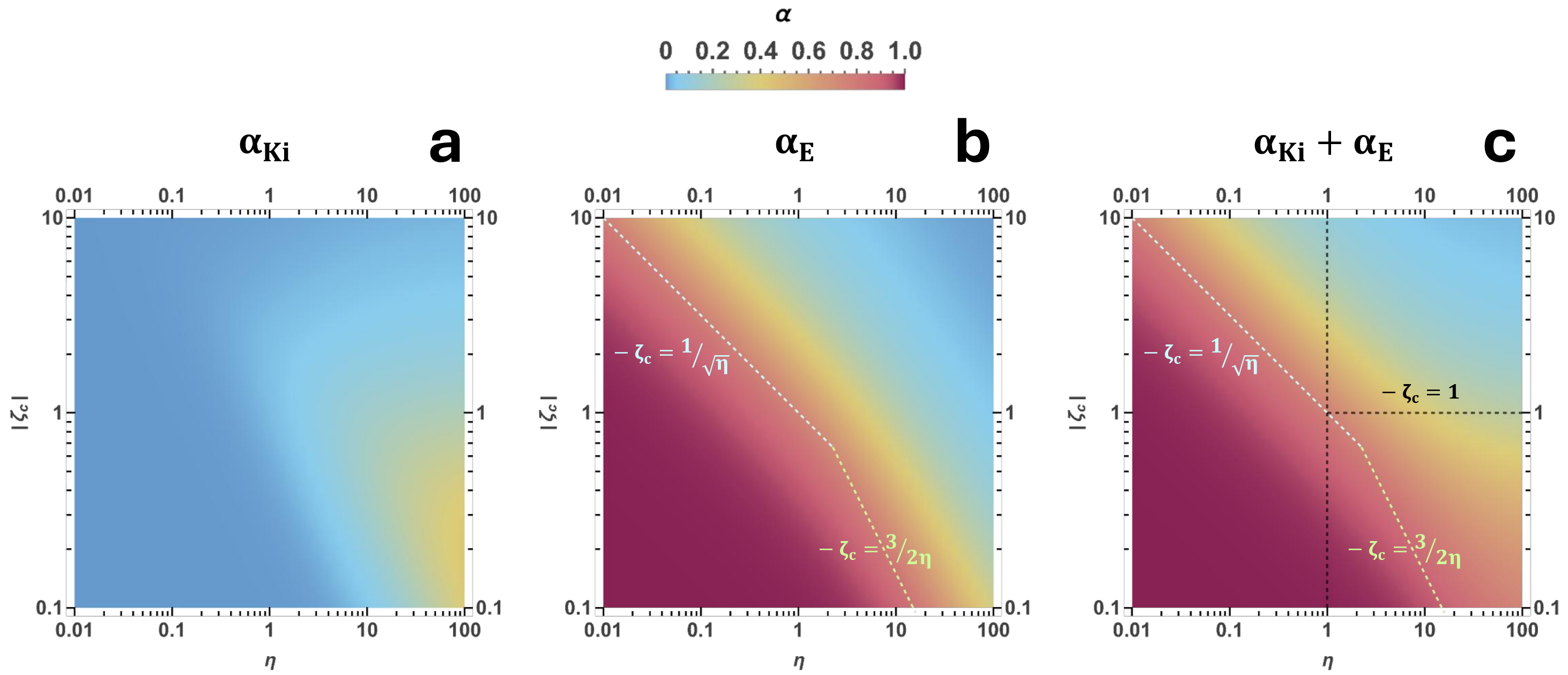}
    \caption{Parametric variations of $(a)$ $\alpha_{Ki}$, $(b)$ $\alpha_E$ and $(c)$ $\alpha_{Ki} + \alpha_E$ for $\eta \in [10^{-2},10^2]$ and $|\zeta_c| \in [0.1,10]$. $-\zeta_c = \sfrac{1}{\sqrt{\eta}}$ for low $\eta$, and $-\zeta_c = \sfrac{3}{2\eta}$ for high $\eta$ are plotted in light blue and light green dashed lines respectively in $(b)$ and $(c)$. In $(c)$, the black dashed lines represent $\eta=1$ and $\zeta_c = -1$}
    \label{fig:alphaPlots}
\end{figure*}

\section{Energetics}\label{sec:energetics}

During the expansion, the thermal energy of the externally heated electrons is converted to electrostatic field energy and thus transferred into kinetic energy of the relatively cold ions. We now discuss the distribution of the total energy $\mathcal{U}_{tot}$ stored in the different channels in the plasma,
\begin{equation}
    \mathcal{U}_{tot} = \mathcal{U}_{Te} + \mathcal{U}_{Ki} + \mathcal{U}_E
\end{equation}
Here, $\mathcal{U}_{Te}$,  $\mathcal{U}_{Ki}$ and $\mathcal{U}_E$ denote the total electron thermal energy, ion kinetic energy and electrostatic energy in the perturbed plasma,
\begin{subequations} \label{EnergyDefs}
        \begin{align}
        \mathcal{U}_{Te} &=  \exp{\left(3\gamma t\right)} \left|\zeta_c\right|\label{UthDef}\\ 
        \mathcal{U}_{Ki} &=  \exp{\left(3\gamma t\right)} \int_{\zeta_c}^{\zeta_f} \frac{1}{2}N_i P_i^2 d\zeta\\
        \mathcal{U}_E &=  \exp{\left(3\gamma t\right)} \frac{1}{\eta} \left[\int_{\zeta_c}^{\zeta_f} \frac{1}{2}Q^2 d\zeta + Q_f \right] \label{EnDefUth}
    \end{align} 
\end{subequations}
where the energies are defined per unit surface, and normalized to $n_{e0}T_{e0} \lambda_{s0}$. The first and second terms on the right hand side of Eq.~\eqref{EnDefUth} are the total electrostatic energies in $\zeta_c\le\zeta\le\zeta_f$ and the electron sheath regions respectively.

To maintain the $T_e$ variation of Eq.~\eqref{genTvar}, the electrons in the expanding self-similar region are assumed to be gaining energy from an external heat source. A part of this energy is distributed to the ions and the electrostatic field in the plasma, which increase $\mathcal{U}_{Ki}$ and $\mathcal{U}_E$. We define partition factors $\alpha_{Ki}$ and $\alpha_E$ for this energy transferred from $\mathcal{U}_{Te}$ to the two channels $\mathcal{U}_{Ki}$ and $\mathcal{U}_E$,
\begin{equation} \label{enPartition}
       \mathcal{U}_{Ki} = \alpha_{Ki} \mathcal{U}_{Te} \qquad ; \qquad \mathcal{U}_E = \alpha_E \mathcal{U}_{Te}
\end{equation}
which upon using Eqs.~\eqref{EnergyDefs} give
\begin{subequations} \label{alphaKiE}
        \begin{align}
        \alpha_{Ki}(\eta,|\zeta_c|) &=  \frac{1}{\left|\zeta_c\right|}\int_{\zeta_c}^{\zeta_f} \frac{1}{2}N_i P_i^2 d\zeta\\
        \alpha_E(\eta,|\zeta_c|) &=  \frac{1}{\eta\left|\zeta_c\right|} \left[\int_{\zeta_c}^{\zeta_f} \frac{1}{2}Q^2 d\zeta + Q_f \right]
    \end{align} 
\end{subequations}
These factors depend on the parameters $\eta $ and $\zeta_c$ and specify the distribution of the total energy among the kinetic energy of ions and electrostatic field energy relative to the total energy in the electrons. The parametric variation of $\alpha_{Ki}$ and $\alpha_E$ for $\eta$ and $\zeta_c$ in the ranges $\left[10^{-2},10^2\right]$ and $\left[0.1,10\right]$ respectively, are shown in figures~\ref{fig:alphaPlots}$(a)$ and  \ref{fig:alphaPlots}$(b)$ respectively. Figure~\ref{fig:alphaPlots}$(c)$ shows the parametric variation of $\alpha_{Ki} + \alpha_E$. This represents the variation of total energy flowing from the electrons to the ions and the field relative to the total electron thermal energy. Since the electrons act as the source of energy for $\mathcal{U}_{Ki}$ and $\mathcal{U}_E$,
\begin{equation} \label{alphaTotCond}
       0 < \alpha_{Ki} + \alpha_E < 1 
\end{equation}
as evidenced in Fig.~\ref{fig:alphaPlots}$(c)$.

In Fig.~\ref{fig:alphaPlots}$(a)$, $\alpha_{Ki}$ increases with $\eta$  at fixed $\zeta_c$ as the ions respond faster and are accelerated to higher energies in the electron heating timescale. Since the strength of the sheath field governs the velocity scale of the accelerated ions, the parametric variation of $Q_f$ (Fig.~\ref{fig:paramVarPlots}$(b)$) determines the variation of the ion kinetic energy with respect to $\eta$ and $|\zeta_c|$. Thus, at low values of $|\zeta_c|$, the total ion kinetic energy increases relative to the total electron thermal energy with increasing $|\zeta_c|$ for fixed $\eta$. However, $\alpha_{Ki}$ gets smaller at large values of $|\zeta_c|$, as the variations in the ion profiles start to become insensitive to $\zeta_c$ while the electron thermal energy increases with increasing $|\zeta_c|$.

The parametric variations of $\alpha_{Ki}$ and the maximum ion kinetic energy $\mathcal{K}_{\rm max}$ are of practical interest in many laser-plasma interaction scenarios, such as for optimizing ion acceleration schemes. $\alpha_{Ki}$ denotes the efficiency with which electron thermal energy is transferred to the ions. $\mathcal{K}_{\rm max}$ can be obtained from $Q_f$ using,
\begin{equation}
         \mathcal{K}_{\rm max}(t) = Z T_{e0} \exp{\left(2\gamma t\right)} \frac{1}{2}P_f^2 = Z T_e(t) \frac{1}{2}Q_f^2
\end{equation}
Since $P_f = Q_f$, the parametric variation of $Q_f$ in Fig.~\ref{fig:paramVarPlots}$(b)$ also indicates the variation of the maximum energy of the accelerated ions.

For low values of $\eta$ and $|\zeta_c|$, the rapid electron heating creates large electron-ion charge separation as observed in regimes $(II)$ and $(III)$ in the Secs.~\ref{sec:lowEta} and \ref{sec:highEta}. With increasing $\eta$ and $|\zeta_c|$ the charge separation, and consequently the total electrostatic energy relative to the thermal energy in the plasma $\left(\alpha_E\right)$, decreases, as observed in Fig.~\ref{fig:alphaPlots}$(b)$. As the plasma profiles saturate for
\begin{equation} \label{DeltaSat}
    |\zeta_c|\gg\Delta\zeta_{\rm sat} = \begin{cases}
        \frac{1}{\sqrt{\eta}} & \eta<1 \\
        1 & \eta>1
    \end{cases}
\end{equation}
as noted in Sec.~\ref{sec:Qual}, so do $\mathcal{U}_{Ki}$ and $\mathcal{U}_E$. But with $\mathcal{U}_{Te}$ increasing proportionally to $|\zeta_c|$, $\alpha_{Ki} + \alpha_E$ asymptotically decreases as ${|\zeta_c|}^{-1}$ for $|\zeta_c|\gg\Delta\zeta_{\rm sat}$. Thus, $\alpha_{Ki} + \alpha_E \rightarrow 0$ for $|\zeta_c|\gg\Delta\zeta_{\rm sat}$ as seen in Fig.~\ref{fig:alphaPlots}$(c)$. Lastly, $\alpha_{Ki} + \alpha_E \rightarrow 1$ in regimes~$(II)$ and $(III)$, i.e., for
\begin{equation}
    |\zeta_c|\ll \begin{cases}
        \frac{1}{\sqrt{\eta}} & \eta \ll 1 \\
        \frac{3}{2\eta}& \eta \gg 1
    \end{cases}
\end{equation}
which is also illustrated in Fig.~\ref{fig:alphaPlots}$(c)$.

\section{Application in laser-plasma interactions}\label{sec:application}
A high-intensity laser pulse generates a plasma at a target surface while the rising intensity profile of the pulse envelope before the pulse peak interacts with the target. The plasma generation begins when the laser intensity becomes high enough to deliver the energy required to ionize the matter at the surface of interest. For instance, if the surface interacting with the laser is considered, field ionization rapidly ionizes the valence shell electrons at this surface, with typical ionization energies in the range of a few $\rm eV$s to $10$s of $\rm eV$s, as the intensity reaches $\rm \sim 10^{14} ~\sfrac{W}{{cm}^2}$. Following ionization, the electrons continue being heated by the rising laser envelope, while the electrostatic field generated by charge separation expands the plasma. The above intensities being relatively weak in the context of intense laser-plasma interactions (normalized vector potential $a\ll1$), the hot electrons remain non-relativistic until $a\lesssim 1$. Thus, the proposed self-similar framework can be used to analyze the expansion dynamics of the plasma at a target surface of interest as long as the electrons at the surface are non-relativistic (i.e. intensities $\rm \lesssim 10^{17} - 10^{18} ~ \sfrac{W}{{cm}^2}$).

In this section, the limiting self-similar solutions are used to estimate the surface plasma dynamics during the interaction of a target with the rising envelope of high intensity laser pulses. The parametric dependencies discovered from these solutions are related to the laser and target parameters. The target surface is ionized to an electron density $n_{e0}$, and the generated plasma expands while being heated by the increasing laser intensity. We choose $t=0$ to denote the onset of collisionless expansion of the plasma, when the electrons have a temperature $T_{e0}$ such that Eq.~\eqref{colLessApprox} holds. The rising intensity envelope is assumed to have a temporal variation of the form
\begin{equation} \label{prepulse}
        \mathcal{I}(t) = \mathcal{I}_i \exp{\left(\frac{2t}{\tau_p}\right)}
\end{equation}
where $\mathcal{I}_i$ is the laser intensity at $t=0$, and $\tau_p$ governs the rise time from the pedestal up to the peak region in the pulse. This form of temporal variation is chosen to coincide with the intensity profile of a $\mathrm{sech}^2$ laser pulse of pulse duration $\tau_p$ far from its peak.

The laser's influence on the plasma expansion is characterized by the energy deposited by it to the electrons. The microphysics of typical laser heating mechanisms depend on the instantaneous intensity of the laser pulse. For simplicity, we introduce an absorption factor $f$ to account for the energy flux from the laser into the participating plasma (directly or indirectly), at the target surface of interest. $f$ is used to relate the incident electromagnetic energy flux to the rates of increases of $\mathcal{U}_{Te}$, $\mathcal{U}_{Ki}$ and $\mathcal{U}_E$ in this plasma,
\begin{equation} \label{En_rates}
        \dfrac{ d\mathcal{U}_{tot}}{dt} = n_{e0}T_{e0}\mkern-3mu\left(\mkern-4mu\frac{C_{s0}}{\gamma}\mkern-4mu\right)\mkern-2mu \frac{d}{dt}\left(\mathcal{U}_{Te} + \mathcal{U}_{Ki} + \mathcal{U}_E\right) = f\mathcal{I}
\end{equation}
Using Eq.~\eqref{enPartition}, Eq.~\eqref{En_rates} can be rewritten as
\begin{equation}
        n_{e0}T_{e0}
\left(\frac{C_{s0}}{\gamma}\right)\frac{d}{dt}\mathcal{U}_{Te} = \frac{f}{\left(1 + \alpha_{Ki} + \alpha_E\right)}\mathcal{I}
\end{equation}
Then, the evolutions of the incident energy flux and the electron thermal energy in the heated plasma can be related using Eqs.~\eqref{prepulse} and \eqref{UthDef},
\begin{equation} \label{thEnLasFluxRel}
        n_{e0}T_{e0}
L_0\exp{\!\left(3\gamma t \right)}= \frac{f\mathcal{I}_i}{3\gamma\left(1 \!+\! \alpha_{Ki} \!+\! \alpha_E\right)} \exp{\mkern-4mu \left(\!\frac{2t}{\tau_p}\!\right)}
\end{equation}
The absorption factor $f$ depends on the mechanisms of laser absorption, plasma formation, and/or energy transfer to the target surface of interest. The limiting self-similar solutions admit temporal variations of $f$ of the form
\begin{equation} \label{f_f0Beta}
        f = f_0 \exp{\left(\beta t \right)}
\end{equation}
for real $\beta$.

Equations~\eqref{thEnLasFluxRel} and \eqref{f_f0Beta} relate the thermal energy deposition rate $3 \gamma$ and length of the heated plasma domain $L=L_0\exp(\gamma t)$ of the limiting self-similar solutions to the laser and plasma parameters and $f$,
\begin{subequations} \label{gamma-L}
        \begin{align}
        3 \gamma &= \frac{2}{\tau_p}+\beta \label{gamma-Laser} \\
        L_0 &= \frac{\tilde{f}_0}{n_{e0}T_{e0}\left(\frac{2}{\tau_p}+\beta\right)}\mathcal{I}_i \label{L0-Laser}
    \end{align} 
\end{subequations}
where $\tilde{f}_0 = f_0 \left(1 + \alpha_{Ki} + \alpha_E\right)^{-1}$. These relations determine the parameters $\eta$ and $|\zeta_c|$, and consequently the expansion phenomenology. The temporal evolution of the length scales Eq.~\eqref{scalesEvol} can be related to the temporal evolution of the envelope intensity Eq.~\eqref{prepulse}, and $\mathcal{I}_i$ using Eq.~\eqref{gamma-Laser} as
\begin{equation} \label{lengths-Laser}
    \frac{L(t)}{L_0} = \frac{\lambda_D(t)}{\lambda_{D0}} = \frac{\lambda_s(t)}{\lambda_{s0}} = \left(\frac{\mathcal{I}(t)}{\mathcal{I}_i}\right)^{\frac{1}{3}\left(1+\frac{\beta}{2}\tau_p\right)}
\end{equation}
and, $L_0$ and $\gamma$ obtained from Eqs.~\eqref{gamma-L} can be used to calculate $\lambda_{D0}$ and $\lambda_{s0}$. These length scales depend on the plasma characteristics at $t = 0$ (when the laser intensity is $\mathcal{I}_i$), since when the collisionless interaction is considered, and can be used to obtain the following expressions for $\eta$ and $\zeta_c$,
\begin{subequations}
        \begin{align}
        \sqrt{\eta} &= 3 \omega_{pi0}\left(\frac{2}{\tau_p}+\beta\right)^{-1} \\
        \left|\zeta_c\right| &= \frac{\tilde{f}_0}{3n_{e0}T_{e0} C_{s0}}\mathcal{I}_i
    \end{align} 
\end{subequations}
These can be rewritten approximately as
\begin{subequations} \label{EtaZetaLP}
        \begin{align}
        \mkern-4mu \sqrt{\eta} &\simeq 0.2 \left(\!\frac{Z^2}{A}\frac{n_{i0}}{\left[10^{22} {cm}^{-3}\right]}\!\right)^{\mkern-4mu\tfrac{1}{2}} \mkern-4mu \frac{\tau_p}{\left[fs\right]} \mkern-2mu \left(1+\frac{\beta \tau_p}{2}\right)^{\mkern-8mu -1} \label{etaEstimate}\\
        \mkern-4mu \left|\zeta_c\right| &\simeq 2\tilde{f}_0 A^{\sfrac{1}{2}} \! \left(\!Z\frac{T_{e0}}{\left[100eV\right]}\!\right)^{\mkern-8mu -\tfrac{3}{2}} \mkern-6mu\left(\!\frac{n_{i0}}{\left[10^{22} {cm}^{-3}\right]}\!\right)^{\mkern-8mu -1} \mkern-15mu \frac{\mathcal{I}_i}{\left[10^{14}W {cm}^{-2}\right]} \label{zetaEstimate}
    \end{align} 
\end{subequations}
where $A$ is the mass number of the ions and $Z$ is their average charge state.

Since $\tilde{f}_0$ in Eq.~\eqref{zetaEstimate} depends on $\alpha_{Ki}+\alpha_E$, it is a function of $\eta$ and $|\zeta_c|$, i.e. $\tilde{f}_0=\tilde{f}_0(f_0,\beta,\eta,|\zeta_c|)$, whose value lies between $\sfrac{f_0}{2}$ and $f_0$ due to condition~\eqref{alphaTotCond}, i.e., $\sfrac{f_0}{2}<\tilde{f}_0<f_0$. Following the discussion for the variation of $\alpha_{Ki} + \alpha_E$ in the previous section, $\tilde{f}_0$ can be obtained in the limiting cases
\begin{equation} \label{f0Asymp}
    \tilde{f}_0 \rightarrow
    \begin{cases}
        f_0 & \mkern6mu |\zeta_c|\gg \Delta\zeta_{\rm sat} \\
        \frac{f_0}{2} & \mkern6mu |\zeta_c|\lesssim \sfrac{1}{\!\!\sqrt{\eta}} \mkern4mu , \mkern4mu \eta<1 \\
        \frac{f_0}{2} & \mkern6mu |\zeta_c|\lesssim \sfrac{3}{2\eta} \mkern6mu, \mkern4mu \eta>1 
    \end{cases}
\end{equation}
where, $\Delta \zeta_{\rm sat}$ is given by Eq.~\eqref{DeltaSat}. For arbitrary values of $\eta$ and $|\zeta_c|$, Eq.~\eqref{zetaEstimate} is an implicit equation for $|\zeta_c|$ that must be solved numerically along with Eqs.~\eqref{alphaKiE}.
\begin{figure*} [!hbtp]
    \centering
    \includegraphics[width=1\linewidth]{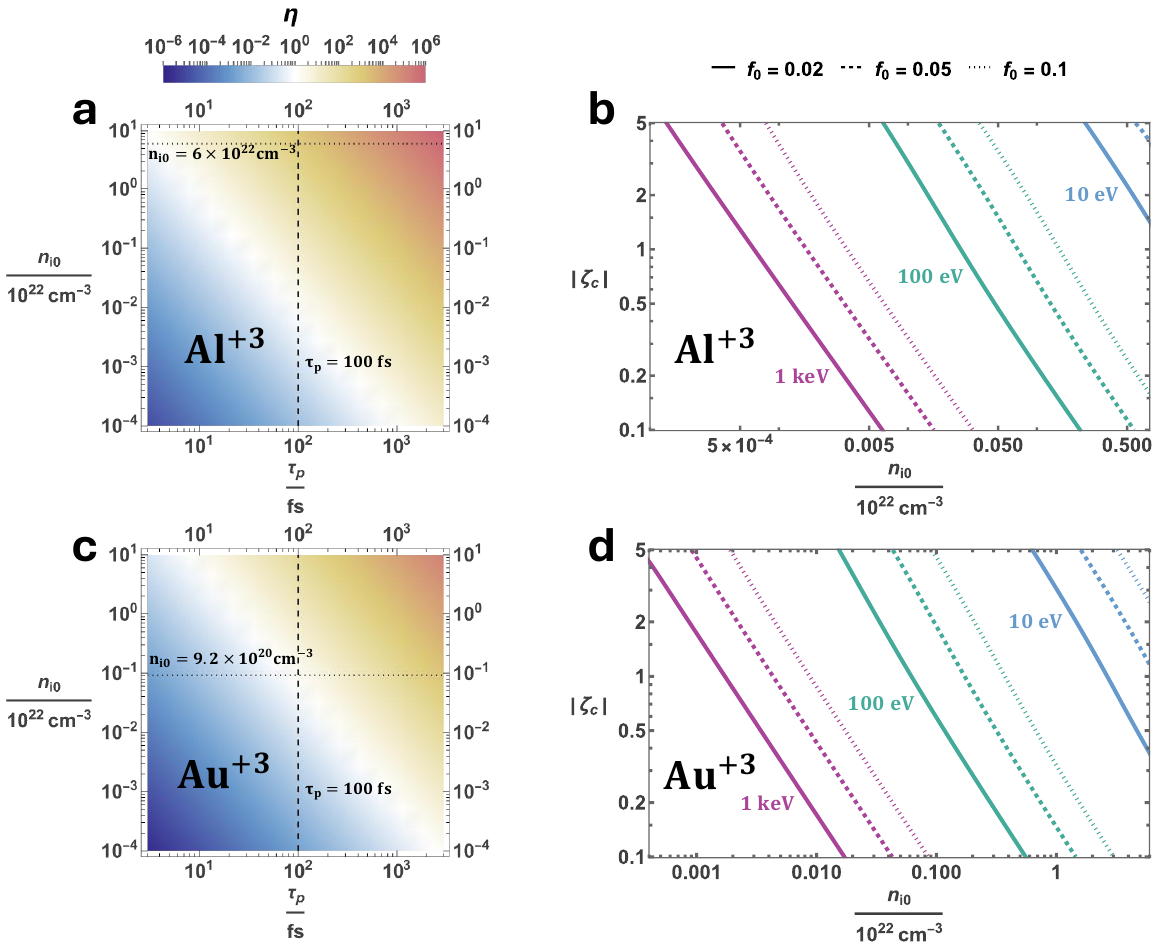}
\caption{Variation of $\eta$ for $n_{i0} \in \left[10^{18}\rm~cm^{-3},10^{23}\rm~cm^{-3}\right]$ and $\tau_p\in \left[3\rm~fs,3\rm~ ps\right]$, and $|\zeta_c|$ with respect to $n_{i0}$ for $f_{0}\in \left\{\rm 0.02, 0.05,0.1\right\}$, for an $\rm Al^{+3}$ plasma $\big(a$ and $b\big)$ and $\rm Au^{+3}$ plasma $\big(c$ and $d\big)$. The $|\zeta_c|$ plots in purple, green and blue for  $T_{e0} \text{ at }{10}^{14} \rm ~\sfrac{W}{cm^2} \in \left\{ 1 ~\rm keV,100 ~eV,10 ~eV\right\}$ respectively are evaluated at $\tau_p=100\rm~ fs$ $\big($denoted by the dashed lines in $(a)$ and $(c)\big)$. Solid density $\rm Al^{+3}$ plasma is denoted by dotted line in $(a)$, while the dotted line in $(c)$ represents $\rm Au^{+3}$ plasma produced from a gold foam target.}
    \label{fig:EtaZetaLP}
\end{figure*}

For an aluminium and a gold target, the dependence of $\eta$ and $|\zeta_c|$ on the laser-plasma interaction parameters given by Eqs.~\eqref{EtaZetaLP} are illustrated in Fig.~\ref{fig:EtaZetaLP}. The ions are assumed to have a mean charge state $Z=3$ and the electron thermal absorption efficiency is assumed to be constant $f_0$ ($\beta=0$) during the interaction. The variation of $\eta$ with respect to the target density $n_{i0}$ in the range $10^{18}\rm ~cm^{-3}$ to $10^{23}\rm~cm^{-3}$, and a pulse rise timescale in the range $3 \rm ~fs$ to $3 \rm~ps$ are shown in plots~\ref{fig:EtaZetaLP}$(a)$ and \ref{fig:EtaZetaLP}$(c)$, for the $\rm Al^{+3}$ and $\rm Au^{+3}$ plasmas respectively. For the $|\zeta_c|$ variation, three different initial temperatures $T_{e0}=\rm \left\{1 ~keV,100~eV, 10~eV\right\}$ in the pulse pedestal at an intensity $\mathcal{I}_i=10^{14}\rm~ \sfrac{W}{cm^2}$ are considered. The variation of $|\zeta_c|$ with the target density $n_{i0}$ for $\tau_p=100 \rm~fs$ are shown in Fig.~\ref{fig:EtaZetaLP}$(b)$ and Fig.~\ref{fig:EtaZetaLP}$(d)$ for the $\rm Al^{+3}$ and $\rm Au^{+3}$ plasmas respectively. Each color denotes a different value for $T_{e0}$, for which absorption factors $f_0 \in \{0.02,0.1,0.5\}$ are plotted with solid, dashed and dotted lines respectively. For the $\alpha_{Ki} + \alpha_E$ dependence of $\tilde{f}_0$ in Eq.~\eqref{zetaEstimate}, the variation plotted in Fig.~\ref{fig:alphaPlots} is used, along with $\eta$ evaluated at a $\tau_p$ of $100 \rm~fs$ using Eq.~\eqref{etaEstimate}. Thus, for plots~\ref{fig:EtaZetaLP}$(b)$ and \ref{fig:EtaZetaLP}$(d)$, the corresponding $\eta$ values lie on the $\tau_p = 100 \rm~fs$ lines shown in plots~\ref{fig:EtaZetaLP}$(a)$ and \ref{fig:EtaZetaLP}$(c)$.

For lower density plasmas and/or for interactions with short pulse rise timescales, the self-similar expansion lies in the low$-\eta$ regimes as noticed in Fig.~\ref{fig:EtaZetaLP}$(a)$ and Fig.~\ref{fig:EtaZetaLP}$(c)$. In terms of plasma composition, $\eta$ is lower for plasmas made from heavier elements and ionized to lower charge states. For example, for an aluminum plasma slab of solid density $n_{i0} = 6 \mkern-2mu\times \mkern-2mu10^{22} \rm~cm^{-3}$ interacting with the pulse with a constant absorption efficiency $f_0$ ($\beta=0$) and a mean charge state $Z=3$ during the interaction, Eq.~\eqref{etaEstimate} gives $\eta\approx0.08 \mkern2mu \left(\frac{\tau_p}{\rm fs}\right)^2$. In this example $\big($represented by dotted line in Fig.~\ref{fig:EtaZetaLP}$(a)\big)$, $\eta>1$ for $\tau_p>3.54 \rm ~fs$, which is true for pulses of interest. On the other hand, consider a gold foam target of density $n_{i0} = 0.092\mkern-2mu\times \mkern-2mu 10^{22} \rm~ cm^{-3}\left(0.3 ~g/cm^3\right)$ with $Z=3$ and $\beta=0$ $\big($dotted line in Fig.~\ref{fig:EtaZetaLP}$(c)\big)$. For the expansion dynamics of this gold target, we obtain $\eta \approx 1.68 \mkern-2mu \times \mkern-2mu 10^{-4} \mkern2mu\left(\tfrac{\tau_p}{fs}\right)^2$, which gives $\eta=1$ for $\tau_p \approx 77 \rm ~fs$. Then both low$-\eta$ and high$-\eta$ regimes can be accessed by varying the pulse rise timescale $\tau_p$ (and $\beta$, when $\beta \neq 0$). For instance, for $\tau_p\approx 24.4 \rm ~fs$ in this example, the $\eta = 0.1$ profiles of Fig.~\ref{fig:paramVarPlots}$(a)$ (in blue) can be obtained, in which the ion profiles are only slightly perturbed from the initial plasma state. However, the $\eta = 100$ profiles in red in Fig.~\ref{fig:paramVarPlots}$(a)$ would be the expected dynamics for a pulse with $\tau_p\approx 0.77 \rm~ps$ interacting with the gold foam target.

As evidenced in figures~\ref{fig:EtaZetaLP}$(b)$ and \ref{fig:EtaZetaLP}$(d)$, for a given plasma composition ($Z$ and $A$), $|\zeta_c|$ decreases with increasing density of target $n_{i0}$, increasing temperature $T_{e0}$, and/or decreasing absorption factor $f_0$. Comparing plots with the same $T_{e0}$ and $f_0$ in Fig.~\ref{fig:EtaZetaLP}$(b)$ with Fig.~\ref{fig:EtaZetaLP}$(d)$, it is observed that $|\zeta_c|$ is higher for $\rm Au^{+3}$ than $\rm Al^{+3}$ with all other parameters the same. In the above two examples of aluminum solid and gold foam targets, with an initial temperature $T_{e0} = 100 \rm~eV$ in the pulse pedestal at an intensity $\mathcal{I}_i=10^{14} \rm~\sfrac{W}{cm^2}$, Eq.~\eqref{zetaEstimate} gives $\zeta_c \approx -0.35 \tilde{f}_0$ and $\zeta_c \approx -62.4 \tilde{f}_0$ respectively. Since $\tilde{f}_0<1$, the dynamics for the first case lies in regimes~$(III)$ or $(IV)$ with $\zeta_c < -0.35$. The asymptotic behavior of regime~$(III)$ is obtained for pulses with $f_0\mkern-2mu\left(\frac{\tau_p}{\rm fs}\right)^2\ll107.1$ (using $\tilde{f}_0\approx \sfrac{f_0}{2}$ due to Eq.~\eqref{f0Asymp}). In the second case, for pulses with $\tau_p\gg77 \rm~fs$ the dynamics lies in regime~$(V)$ when $\tilde{f}_0 \gtrsim 0.016$, and in regime~$(III)$ when $f_0\mkern-2mu\left(\frac{\tau_p}{\rm fs}\right)^2\ll286.2$ using Eq.~\eqref{f0Asymp}. For instance, the solutions plotted in Fig.~\ref{fig:HeHzplots} would be the expected dynamics in this example for absorption factors $f_0 > 0.016$ during the interaction of the gold foam plasma with the laser pulse. A pulse rise time $\tau_p\approx 0.77 \rm~ps$ would produce the $\eta=100$ profiles in Fig.~\ref{fig:HeHzplots}$(a)$ - $(c)$, while the $\eta=900$ dynamics in Fig.~\ref{fig:HeHzplots}$(d)$ - $(f)$ would be obtained for $\tau_p\approx 2.3 \rm~ps$. $\tilde{f}_0$ can be similarly approximated in the low$-\eta$ regimes when $\tau_p<77 \rm~fs$ using the limits provided by Eq.~\eqref{f0Asymp}.

\section{Conclusion}
To summarize, we have formulated a new family of self‐similar equations for collisionless plasma expansion with active heating/cooling. The formulation extends the scope of self-similar analysis in the presence of external electron energy transfer mechanisms that produce $T_e$ variations of the form Eq.~\eqref{genTvar}. For electrons with uniform temperature profiles $T_e(t)$, Eqs.~\eqref{SSuniGen:sys} - \eqref{scaledSSvarGen} constitute a three-parameter self-similar framework for plasma dynamics governed by the electrostatically coupled two-fluid Eqs.~\eqref{fluidlab:sys}. This can be used to study self-similar plasma expansion for a range of different $T_e(t)$ variations given by Eq.~\eqref{genTvar_uni}.

A key insight of the self-similar solutions is the correlation length $\lambda_s(t) = \sfrac{C_{s0}}{\gamma} + \int_0^t C_s(t') dt'$ that emerges from the ion momentum equation during the self-similar reduction. It characterizes the distance traveled by acoustic perturbations in the expansion timescale, and serves as a natural scale for the self-similar expansion with active heating/cooling. The magnitudes of the Debye length $\lambda_D(t)$ and the length of the heated plasma domain $L(t)$ relative to this length scale determine vastly different qualitative behaviors for the expanding plasma. The relative magnitudes of these three scales are characterized by the parameters $\eta=\left(\sfrac{\lambda_s}{\lambda_D}\right)^2$ and $|\zeta_c|=\sfrac{L}{\lambda_s}$. The plasma behaviors have been classified into five distinct dynamical regimes, ranging from dynamics that lead to subsequent Coulomb explosion to those characterized by nearly quasineutral expansion. The proposed self-similar formulation thus provides a unified framework to describe the expansion dynamics. We have demonstrated the continuous transition among the regimes in the $\eta-|\zeta_c|$ parameter space for the case of self-similar solutions with an exponentially varying uniform electron temperature $T_e(t)=T_{e0}\exp{\left(2 \gamma t\right)}$, and detailed the asymptotic dynamics in the regimes.

In the low$-\eta$ regimes, rapid electron heating drives an early formation of an electron sheath and electron dynamics in a boundary layer while the ions remain nearly unperturbed. When $|\zeta_c|$ is large $\big($regime~$(I)\big)$, the sheath field is shielded in a region much smaller than the distance over which the plasma is heated, $\lambda_s\ll\lambda_D \ll L$. This behavior is particularly useful in applications where a modest or limited ion expansion is desired - for instance, in target pre-expansion scenarios where the integrity of the target must be preserved, like in laser-plasma fusion contexts. The dynamics in regime~$(II)$ yields an almost bare, unperturbed ion slab ($\omega_{pi0}\ll\gamma$ or $\lambda_s\ll\lambda_D$) following electron evacuation from the small domain of  heating with $L \lesssim \lambda_D$. Such conditions can trigger a subsequent Coulomb explosion of the ion slab, which can be exploited in applications where rapid ion acceleration from a surface is desired after the plasma is heated. The predictions in this regime could also be applicable in understanding disintegration of laser-irradiated nanostructured targets. In the context of laser-plasma interactions, targets of low density, made from heavier elements, and/or with high absorption efficiency would be ideal for regime~$(I)$ dynamics. On the other hand, low density targets with low absorption efficiency would be required for regime~$(II)$.

For the high$-\eta$ regimes, slower electron heating relative to the ion response leads to more pronounced ion dynamics. In regime~$(III)$, where $L \ll \lambda_D \ll \lambda_s$, a thin ion slab with a non-uniform density gradient is produced from the small region of the heated surface plasma. The ions in the slab are not strongly accelerated and maintain subsonic velocities during the heated expansion. Operating in this regime during laser-target interactions could be useful for controlled surface modification of the target. For instance, the predictions in this regime can guide experimental design for laser-plasma schemes where a strong density gradient is desired at the target surface with modest ion energies before the pulse peak interacts. $\lambda_D \ll \lambda_s$ in regimes~$(IV)$ and $(V)$ leads to strong modulations in the electric field and density profiles, and high maximum kinetic energies of the ions with supersonic velocities. Since, the heating domain is small in regime~$(IV)$, $L\ll\lambda_s$, the conversion efficiency of electron thermal energy to the total ion kinetic energy is highest in this regime, and may be particularly advantageous for laser-driven ion acceleration schemes, and for generating high-energy, quasi-monoenergetic ion beams. Lastly, the nearly quasineutral expansion in regime~$(V)$ with a large heating domain $L\gtrsim \lambda_s$, is useful for understanding plasma expansion in laser-plasma schemes with bulk heating, or for sustained plasma expansion scenarios encountered in astrophysical scenarios. In the context of laser plasma interaction, the high$-\eta$ dynamics are obtained for slow electron heating by the laser relative to ion response ($\gamma \ll \omega_{pi0}$). High density targets made of lighter elements are desirable for dynamics in regimes~$(III)$ and $(IV)$, such that $\eta\gg1$ and $|\zeta_c|\ll1$ in Eq.~\eqref{EtaZetaLP}. While low absorption efficiency $f_0$ produces dynamics in these low$-|\zeta_c|$ regimes, target plasmas with high $f_0$ could lead to the high$-|\zeta_c|$ dynamics of regime~$(V)$.

\setlength{\tabcolsep}{12pt}
\setlength{\cellspacetoplimit}{2pt} 
\setlength{\cellspacebottomlimit}{2pt} 
\begin{table*} [!hbtp]
    \centering
    \begin{tabular}{Sc Sc Sc Sc Sc Sc}\toprule
         &  \textbf{Regime $(I)$}&  \textbf{Regime $(II)$}&  \textbf{Regime $(III)$}&  \textbf{Regime $(IV)$}& \textbf{Regime $(V)$}\\
         \midrule
         \textbf{Ordering}&  $L\gg\lambda_D \gg \lambda_s$&  $L\lesssim \lambda_D \quad \lambda_D \gg \lambda_s$&  $L \ll \lambda_D \ll \lambda_s$&  $L \ll \lambda_s \quad \lambda_D \ll \lambda_s$& $L \gtrsim \lambda_s \gg \lambda_D$\\         
         \midrule
         $\mathbf{\Delta L_e}$&  $\lambda_D$&  $\sfrac{\lambda_D^2}{L}$&  $\sfrac{\lambda_D^2}{L}$&  $\lambda_s$& $\lambda_s$\\ 
         $\mathbf{ \Delta L_i}$&  $\sfrac{\lambda_{s}^2}{\lambda_{D}}$&  $\sfrac{\lambda_s^2 L}{\lambda_D^2}$&  $L$&  $L$& $\lambda_s$\\
         $\mathbf{ \Delta L_{i,\rm drop}}$&  &  &  $\sfrac{\lambda_DL}{\lambda_s}$&  $\sfrac{\lambda_DL}{\lambda_s}$& \\ 
         \bottomrule
    \end{tabular}
    \caption{Orderings of the length scales $L$, $\lambda_D$ and $\lambda_s$ in the different regimes, and dependence of the variation scales of the electrons and ion fluids on these characteristic scales. These variation scales are assigned based on the representative shapes of the electron and ion density profiles in each regime. In regimes~$(III)$ and $(IV)$, the ion profile drops steeply near the unperturbed plasma interface on a scale $\Delta L_{i,\rm drop}\ll \Delta L_i$}
    \label{tab:LengthScales}
\end{table*}

In the solutions where $\eta \gtrsim 1$, the ion density profiles near the ion expansion front $\zeta_f$ exhibit oscillations about the electron density $N_{ef}$ on a very small scale. These features complemented by the flattening of the ion velocity profiles near $\zeta_f$ are expected to result in peaks in the ion kinetic energy spectrum at high energies. Analysis of these hydrodynamic shock like features, and of the ion energy spectrum of the expanding plasma will be detailed in a future article.

\begin{figure}
    \centering
    \includegraphics[width=1\linewidth]{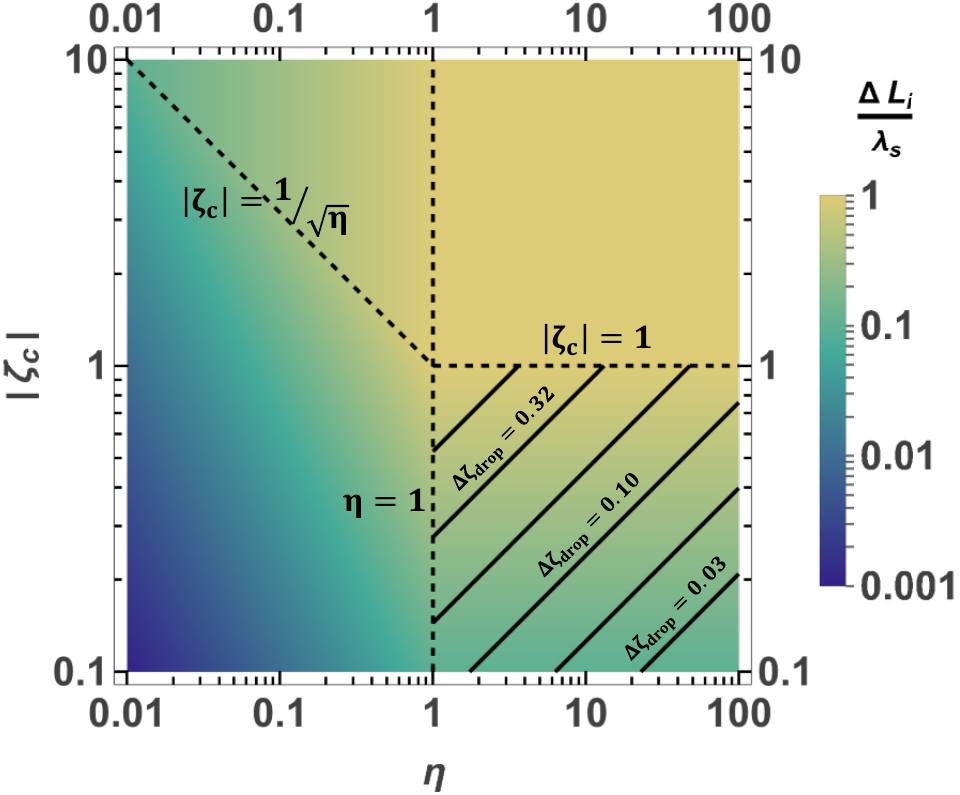}
    \caption{Parametric variation of $\frac{\Delta L_i}{\lambda_s}$ with respect to $\eta\in\left[0.01,100\right]$ and $|\zeta_c|\in \left[0.1,10\right]$. Some contours of $\Delta \zeta_{\rm drop}=\frac{\Delta L_{i,\rm drop}}{\lambda_s}$ in regimes $(III)$ and $(IV)$ are shown. The black dashed lines represent the boundaries $\eta=1$, $|\zeta_c|=\sfrac{1}{\sqrt{\eta}}$ for $\eta<1$, and $|\zeta_c|=1$ for $\eta>1$}
    \label{fig:deltaL}
\end{figure}

Apart from the ion energy variation discussed in Sec.~\ref{sec:energetics}, another relevant property for many applications is the length scale of variation of the electron and ion profiles in the expanding plasma. These variation scales depend on the expansion regime and hence, on the characteristic length scales $L$, $\lambda_D$ and $\lambda_s$. This dependence of the variation scales $\Delta L_i$ and $\Delta L_e$ for the five asymptotic expansion regimes are summarized in Table~\eqref{tab:LengthScales}, along with the orderings of the characteristic length scales in each regime. In regime~$(I)$, the electron density varies over a length scale $\Delta L_e = \lambda_D$, as suggested by Eq.~\eqref{regI_DLe}. For dynamics in regime~$(II)$, the electrons from the unperturbed ion slab of dimension $L$ are blown out to a density $N_e \approx \sfrac{\eta \zeta_c^2}{2}$. The electron fluid in regime~$(III)$ also drops to a low value of $N_e \approx \sfrac{\eta \zeta_c^2}{2}$ in the expanding ion region of length $L_i=\left(1 + \frac{\lambda_s^2}{\lambda_D^2}\right) L$. The ion profiles in this region vary on a length scale $\Delta L_i=L$, while dropping steeply near $x=0$ on a scale $\Delta L_{i,\rm drop} = \frac{\lambda_DL}{\lambda_s}$ as given by Eq.~\eqref{RegIIIDLidrop}. In regime~$(IV)$, the electron density remains small ($\mathcal{O}(|\zeta_c|)$), and the ion density profiles vary on a length scale $\Delta L_i=L$ with $\Delta L_{i,\rm drop} = \frac{\lambda_DL}{\lambda_s}$ near the initial vacuum plasma interface similar to regime~$(III)$. This sharp drop in the ion density, is followed by a low density quasineutral tail with density decreasing from $\mathcal{O}(|\zeta_c|)$ near $x=0$ to  $\mathcal{O}(\sfrac{1}{\eta})$ near $x_f\sim\lambda_s$. The approximately quasineutral dynamics in regime~$(V)$, has a length scale of $\lambda_s$ (Eq.~\eqref{regV_DL}). The electron sheath beyond the ion expansion front in all the regimes has a length scale of $\sfrac{2\lambda_s}{Q_f}$. The dependence of $\frac{\Delta L_i}{\lambda_s}$ on $\eta$ and $|\zeta_c|$ is shown through the color density variation in Fig.~\ref{fig:deltaL}. Some contours of $\Delta \zeta_{\rm drop}=\frac{\Delta L_{i,\rm drop}}{\lambda_s}$ in regimes~$(III)$ and $(IV)$ are also plotted.

We provide several testable signatures of the heated plasma dynamics, including the characteristic length and energy scales, characteristic density and field profiles, and regime transitions in these quantities. The dependence of these quantities on the laser and target parameters can serve as practical guides for designing high intensity laser-plasma experiments. The framework can further be used to analyze the effects of realistic heating mechanisms on the nature of plasma expansion, by coupling the self-similar system derived in Sec.~\ref{sec:setup} with an equation for the rate of electron heating. Equations~\eqref{SSanz:denVel} - \eqref{fluidSS:gen} when supplemented with the appropriate equation for $\Theta(\xi)$, facilitate studying scenarios where non-uniform spatial variations of the electron temperature might be applicable. Possible next steps include calculating the ion energy spectra and comparing our solutions to PIC simulations. Other routes for development include extension of the model to finite sized plasmas in arbitrary geometries, and analysis of the self-similar shock structure predicted by the dispersive hydrodynamic system.

\section*{Acknowledgments}
R.S. received support from TAU Systems under Sponsored Research Agreement UTAUS-FA00001488. This material is based upon work supported by the U.S. Department of Energy, National Nuclear Security Administration under Award Number DE-NA0004201, by the National Science Foundation under Grant Number NSF2108921; and by the Air Force Office of Scientific Research under Award Number FA9550-25-1-0286.

\bibliographystyle{unsrt}
 \bibliography{References}

\onecolumngrid
\appendix

\section{Ion Plasma profiles near the density drop in regimes~$(III)$ and $(IV)$}\label{App:Prof34}
In a region of $\mathcal{O}(|\zeta_c|)$ near the origin in regimes~$(III)$ and $(IV)$ the ions exhibit a steep drop in their density profiles, where the electron-ion charge separation is high, $N_e \ll N_i$. Then, the flow in this region can be modeled using
\begin{subequations}
    \begin{align}
       & N_i\dfrac{dP_i}{d\zeta} + \left(P_i - \zeta_c - \chi\right)\dfrac{dN_i}{d\zeta } = 0 \\
       & P_i - Q + \left(P_i - \zeta_c - \chi\right)\dfrac{dP_i}{d\zeta } = 0 \\
       & \dfrac{dQ}{d\zeta}=\eta N_{i}
    \end{align}
\end{subequations}
with $\chi = \zeta - \zeta_c$. Assuming the ion fluid profiles vary on a scale much smaller than $\chi$, 
\begin{equation} \label{near0_ass}
     \frac{1}{\chi} \ll \left|\dfrac{d\ln{(G_i)}}{d \zeta}\right| \qquad , \qquad \text{where } \mkern 2mu G = \{N,P\} ~,
\end{equation}
and using Eqs.~\eqref{coreBCs}, we obtain the relations for $N_i$ and $Q$ in terms of $P_i$
\begin{subequations}
    \begin{align}
        N_i &= \frac{\left|\zeta_c\right|}{P_i-\zeta_c - \chi} \\
        Q &= P_i + \eta \left|\zeta_c\right| \left(1 + W\left[-\exp{\left(-1-\sfrac{P_i}{\eta\left|\zeta_c\right|}\right)}\right]\right) \label{near0_3,4:Q-P}
    \end{align}
\end{subequations}
where, the Lambert $W$ function is the inverse of the function $f(w) = w e^w$. The ion velocity $P_i$ can be obtained by solving the differential equation
\begin{equation} \label{near0_3,4:dPdZ}
    \dfrac{dP_i}{d \zeta} = \frac{\eta \left|\zeta_c\right| \left(1 + W\left[-\exp{\left(-1-\sfrac{P_i}{\eta\left|\zeta_c\right|}\right)}\right]\right)}{P_i -\zeta_c - \chi}
\end{equation}

For $\sqrt{\sfrac{P_i}{\eta\left|\zeta_c\right|}}\ll 1$, the Lambert $W$ function in Eqs.~\eqref{near0_3,4:Q-P} and \eqref{near0_3,4:dPdZ} can be expanded around $\sfrac{-1}{e}$ to yield,
\begin{equation} \label{W_approx}
     W\left[-\exp{\left(-1-\sfrac{P_i}{\eta\left|\zeta_c\right|}\right)}\right] \approx -1 + \sqrt{\sfrac{2 P_i}{\eta\left|\zeta_c\right|}} + \mathcal{O}\left(\sfrac{P_i}{\eta\left|\zeta_c\right|}\right)
\end{equation}
Then the ion profiles to leading order in $\mathcal{O}\left(\sfrac{P_i}{\eta\left|\zeta_c\right|}\right)$ can be approximated by
\begin{subequations}
    \begin{align}
        N_i &= \left|\zeta_c\right| \left(|\zeta|\left(B^{\frac{1}{3}} + \sgn(\zeta) B^{-\frac{1}{3}}\right)^2 - \zeta\right)^{-1}\\
        P_i &= |\zeta|\left(B^{\frac{1}{3}} + \sgn(\zeta) B^{-\frac{1}{3}}\right)^2
    \end{align}
\end{subequations}
where $\sgn$ is the sign function, and
\begin{equation} \label{BVal}
    B = \sqrt{\frac{9 \eta}{8} \left|\frac{\zeta_c}{\zeta}\right|\left(\sgn(\zeta) + \left|\frac{\zeta_c}{\zeta}\right|\right)^2 - \sgn(\zeta)}
    + \sqrt{\frac{9 \eta}{8} \left|\frac{\zeta_c}{\zeta}\right|}\left(\sgn(\zeta) + \left|\frac{\zeta_c}{\zeta}\right|\right)
\end{equation}

\section{2-scale expansion for the electrostatic waves in regime~$(V)$}\label{regV2scale}
To obtain the electrostatic waves in the bulk of the rarefaction wave in regime~$(V)$, we carry out a two-scale expansion of the system by introducing a small scale variable $\tilde{\zeta}=\zeta/\epsilon$ with $\epsilon \sim \mathcal{O}\left(\sfrac{1}{\sqrt{\eta}}\right)$. The quantities are expanded around the mean solutions
\begin{subequations} \label{He_Hz_Ansatz}
        \begin{align}
        N_i^{(V)} &= \bar{N}(\zeta) +  \boldsymbol{N_i}^{(V)}(\zeta,\tilde{\zeta}) \\
        P_i^{(V)} &= \bar{P}_i(\zeta) +  \boldsymbol{P_i}^{(V)}(\zeta,\tilde{\zeta}) \\
        N_e^{(V)} &= \bar{N}(\zeta) +  \boldsymbol{N_e}^{(V)}(\zeta,\tilde{\zeta}) \\
        Q^{(V)} &= \bar{Q}(\zeta) +  \boldsymbol{Q}^{(V)}(\zeta,\tilde{\zeta})
    \end{align} 
\end{subequations}
where $\boldsymbol{P_i}^{(V)},\boldsymbol{N_\alpha}^{(V)} = \mathcal{O}(\sfrac{1}{\eta})$ and $\boldsymbol{Q}^{(V)} = \mathcal{O}(\sfrac{1}{\sqrt{\eta}})$. Then the oscillatory behavior in this region is governed by the equations
\begin{subequations} \label{He_Hz_osc}
        \begin{align}
        \frac{\partial\boldsymbol{N_i}^{(V)}}{\partial\tilde{\zeta}} &= -\frac{\bar{N}}{\left(\bar{P}_i  -\zeta\right)}\frac{\partial \boldsymbol{P_i}^{(V)}}{\partial\tilde{\zeta}} \\
        \frac{\partial \boldsymbol{P_i}^{(V)}}{\partial\tilde{\zeta}} &= \frac{1}{\left(\bar{P}_i  -\zeta\right)}\boldsymbol{Q}^{(V)} \\
        \frac{\partial\boldsymbol{N_{e}}^{(V)}}{\partial\tilde{\zeta}} &= -\bar{N} \boldsymbol{Q}^{(V)} \\
        \frac{\partial\boldsymbol{Q}^{(V)}}{\partial\tilde{\zeta}} &= \eta (\boldsymbol{N_i}^{(V)}-\boldsymbol{N_e}^{(V)}) -  \frac{d\bar{Q}}{d\zeta}
    \end{align} 
\end{subequations}
The perturbed quantities take the form $\boldsymbol{N_\alpha}^{(V)}(\zeta,\tilde{\zeta}) = \boldsymbol{\overline{N}_\alpha}(\zeta) + \boldsymbol{\widetilde{N}_\alpha}(\zeta,\tilde{\zeta})$, $\boldsymbol{P_i}^{(V)}(\zeta,\tilde{\zeta}) = {\boldsymbol{\overline{P}_i}}(\zeta) + \boldsymbol{\widetilde{P_i}}(\zeta,\tilde{\zeta})$ and $\boldsymbol{Q}^{(V)}(\zeta,\tilde{\zeta}) = \widetilde{\boldsymbol{Q_{}}}(\zeta,\tilde{\zeta})$ where $\boldsymbol{\overline{P}_i}(\zeta)$, $\boldsymbol{\overline{N}_\alpha}(\zeta)$ are perturbations to the mean behaviour on the order $\mathcal{O}(\sfrac{1}{\eta})$. The superposed oscillations given by the solutions of Eqs.~\eqref{He_Hz_osc} take the form
\begin{subequations} \label{He_Hz_Sols}
        \begin{align}
        \boldsymbol{\widetilde{N}_i} &= \frac{\bar{N}}{k(\bar{P}-\zeta)^2}\boldsymbol{A}\exp{\left(ik\zeta\right)} \\
       \boldsymbol{\widetilde{P_i}} &= \frac{-1}{k(\bar{P}-\zeta)}\boldsymbol{A}\exp{\left(ik\zeta\right)} \\
        \boldsymbol{\widetilde{N}_e} &= \frac{\bar{N}}{k}\boldsymbol{A}\exp{\left(ik\zeta\right)} \\
       \widetilde{\boldsymbol{Q_{}}} &= -i\boldsymbol{A}\exp{\left(ik\zeta\right)} 
    \end{align} 
\end{subequations}
where $k^2(\zeta)=\eta\bar{N}\left(\tfrac{1}{(P-\zeta)^2}-1\right)$, and $\boldsymbol{A}(\zeta)$ is the complex envelope for the field oscillations with $\left|\boldsymbol{A}\right| = \mathcal{O}(\sfrac{1}{\sqrt{\eta}})$. $\boldsymbol{\overline{N}_i} \! - \! \boldsymbol{\overline{N}_e}$ is approximately $\tfrac{1}{\eta}\frac{d\bar{Q}}{d\zeta}$, which forms the mean profile for charge separation $ N_i^{(V)} \mkern-5mu -\mkern-4mu  N_e^{(V)}$ in the expansion bulk.

\end{document}